\documentclass[12pt]{article}
\usepackage[top=1in, bottom=1in, left=1.in, right=1.in]{geometry}
\usepackage[utf8]{inputenc}
\usepackage{authblk} 
\usepackage{amsmath,amssymb,amsbsy,amstext,amsthm,IEEEtrantools}
\usepackage{graphicx,caption,subcaption} 
\usepackage[colorlinks=true, allcolors=blue]{hyperref}
\usepackage{physics}
\usepackage{tikz}
\usepackage[numbers,sort&compress]{natbib}
\usepackage{color}
\usepackage{enumitem}
\usepackage{booktabs}

\newcommand{\Retr}{\operatorname{Re}\Tr}

\begin{document}

\begin{titlepage}

\vskip 2.2cm

\begin{center}

{\large \bf  Impact of Irrep Truncations \\
on Quantum Simulations of QCD}
\vskip 1.4cm

{Patrick Draper\footnote{pdraper@illinois.edu} and Drishti Gupta\footnote{dg36@illinois.edu}}
\\
\vskip 1cm
{Department of Physics, University of Illinois, Urbana, IL 61801}\\
\vspace{0.3cm}
\vskip 4pt

\vskip 1.5cm

\begin{abstract}
Quantum simulations of lattice gauge theories with continuous gauge groups require a choice of basis (how states of the gauge theory map to the qubit computational basis) and truncation scheme (which of the infinitely many gauge theory basis states should be omitted in the map to the finite dimensional qubit Hilbert space.) The most common choices are adapted to an electric basis for the gauge theory, where the basis states for each link are associated with irreducible representations of the gauge group. Electric bases and truncations are generally advantageous for preserving gauge symmetries and simulating the strong coupling regime, but their effectiveness in the weak coupling regimes of physical interest are less well understood. In this work, we compute relative probabilities $p(R')/p(R)$ for finding a link in the irreps $R,R'$ in the ground state of SU(3) lattice gauge theory using Euclidean Monte Carlo methods and weak-coupling lattice perturbation theory. The observables that compute these relative probabilities are constructed by introducing seams into an anisotropic lattice, and by a similar construction we are also able to measure relative probabilities of different site singlets (ways that electric flux braids together into a gauge singlet) at each vertex. With these methods we can quantify, for example, the importance of higher electric flux states in the weak coupling ground state. These results directly inform what truncations are needed to accurately capture nonabelian gauge dynamics near the continuum limit. 
\end{abstract}

\end{center}

\vskip 1.0 cm

\end{titlepage}
\setcounter{footnote}{0} 
\setcounter{page}{1}
\setcounter{section}{0} \setcounter{subsection}{0}
\setcounter{subsubsection}{0}
\setcounter{figure}{0}

\tableofcontents

\newpage 
\section{Introduction}

Simulating the real-time dynamics of non-Abelian gauge theories from first principles is one of the most compelling potential applications of quantum computers~\cite{Feynman1982,Jordan2012,Preskill2018,CREUTZ-LGT-MC-REVIEW}. Classical Monte Carlo methods, despite their remarkable success in computing static and thermodynamic properties of quantum chromodynamics (QCD)~\cite{B-D-DECAY-HYPER-PRECISION,flag-review-2024,Borsanyi2016,Finite-Temp-LQCD-CHiral-Crossover}, are fundamentally obstructed from accessing real-time evolution and finite-density regimes by the sign problem~\cite{Troyer2005}. Quantum simulation offers a route around these obstacles, with the prospect of computing scattering amplitudes, thermalization dynamics, and transport coefficients directly from the underlying $SU(3)$ gauge theory~\cite{Bauer2023,Davoudi2020,Klco2022,halimeh2025quantumsimulationoutofequilibriumdynamics}.

The central technical obstacle in this program is that the physical Hilbert space of a lattice gauge theory is infinite-dimensional: each link carries a quantum degree of freedom valued in the Lie group $SU(3)$. The Hilbert space of square-integrable functions on $SU(3)$  decomposes into an infinite tower of irreducible representations (irreps) in the representation basis~\cite{Kogut1975}. A digital quantum computer, by contrast, can only store a finite number of qubits per link, thereby requiring a cutoff scheme in order to map the link Hilbert space to a logical qubit space~\cite{Somma2003,Byrnes2006}. A physically motivated truncation scheme is therefore indispensable, and the quality of any quantum simulation depends critically on how faithfully the truncated Hilbert space captures the low-energy physics of the continuum limit~\cite{siew2026continuumlimitqubitregularizedsu3,Jakobs2025,kane2025obtainingcontinuumphysicsdynamical,Ciavarella2021,Ciavarella-large-N-truncation,Alexandru-S1080-truncation} and informs the choice of truncation for mapping lattice gauge theories on near term quantum hardware~\cite{GonzalezCuadra2025,xu2026observationglueballexcitationsstring,froland2025simulatingfullygaugefixedsu2,ciavarella2026quantumsimulationstwodimensionalnonabelian,Chernyshev2026,Ciavarella-large-N-truncation,balaji2026timedomainobservablesresonantscattering}. 

The representation/electric basis, in which link states are labeled by $SU(3)$ irrep quantum numbers, admits a hierarchy of physically motivated truncation schemes.  These include truncation by the maximum number of tensor indices, $p+q \leq \Lambda$, where $(p,q)$ is the Dynkin label of the irrep and $\Lambda$ is the truncation order~\cite{Ciavarella2021,quantum-circuits-draper}, truncation by the maximum electric energy permitted on a link, and truncation by the maximum electric energy permitted at a vertex~\cite{B-truncation}. The latter scheme of truncation provides finer control in the size of the Hilbert space and the number of magnetic matrix elements by restricting the number of gauge invariant states (``site-singlets") allowed at a vertex.  The accuracy of these truncations in reproducing the physics of the untruncated theory depends on how rapidly the amplitudes of irrep and site-singlet probabilities fall off in the physical vacuum and low-lying states of the lattice Hamiltonian. 

Raising these truncations certainly improves accuracy. On the other hand, it also makes quantum simulations more resource-intensive. Thus it is desirable to balance cost against fidelity to the untruncated continuum theory when developing  simulations of lattice gauge theories for fault-tolerant quantum devices. Qubit and gate costs for scalable simulations of $U(1)$, $SU(2)$, $SU(3)$, and QCD have been studied extensively for different truncation schemes~\cite{Lamm2024,Rhodes2024,kan2022latticequantumchromodynamicselectrodynamics,draper2026blockencodingnonabelianlattice,Hidalgo2026}. For instance, the works of Rhodes et al~\cite{Rhodes2024} and Kan et al~\cite{kan2022latticequantumchromodynamicselectrodynamics} estimate resources in a $\Lambda$ truncation scheme with a conservative choice of $\Lambda$. $2\Lambda \sim 10$.  However, precise computations of physically motivated target truncations is lacking. A campaign to quantify irrep and site-singlet probability distributions directly from the  path integral could  inform optimal truncation schemes and resource estimates for quantum simulations in the electric basis and its descendants. Here we develop tools to compute these probability distributions and estimate what minimal truncations are sufficient to reproduce low-lying continuum physics. 

The primary tool we develop is a Monte Carlo method that directly measures the relative probability of irreps and site-singlet configurations in the ground state of the $SU(3)$ Kogut-Susskind (KS) Hamiltonian~\cite{Kogut1975}. Our approach exploits the connection between the KS Hamiltonian and the Euclidean path integral in the anisotropic limit~\cite{creutz-transfer-matrix,hasenfratz-hamiltonian-scale,Karsch1982}. Measurements are performed in the representation basis on an anisotropic Euclidean lattice, where the temporal lattice spacing $a_t \ll a_s$, probe the continuum ground-state wavefunction in the infinite volume limit. Conventional path integral methods are defined over a group element measure; in this basis, irrep and site-singlet probabilities correspond to off-diagonal observables. Therefore we must measure them by inserting doubled link variables on measured links, which we will refer to as ``seams".\footnote{This construction is a general technique for computing off-diagonal observables, such as the worm operator in quantum Monte Carlo algorithms~\cite{Worm-Operator}.}

We implement this seamed lattice path integral measurements using two lattice actions, the standard Wilson plaquette action~\cite{Wilson1974} and the heat kernel action~\cite{Menotti1981288,Drouffe1983}, across a range of bare couplings $\beta$ on a $4$-dimensional anisotropic lattice with anisotropy $\xi = a_s/a_t$. Measurements of link irrep probabilities for coupling $\beta=6.0$ indicate that irreps containing two-index and three-index tensors have non-trivial abundance in the ground state. Furthermore, measurements of site-singlet probabilities indicate sizable abundance of singlets formed by the fundamental and anti-fundamental irreps; however, measurements of singlet probabilities involving higher irreps are mostly noise-limited. We then compare values obtained from Monte Carlo to tadpole-improved perturbation theory~\cite{Lepage-tadpole} using the link-field renormalization factors $u_s$ and $u_t$ extracted from measured plaquette expectation values.

The work is structured as follows: In Sec.\,\ref{sec:weak-pt}, we will derive perturbative expressions for electric correlators, which provide intuition and useful comparisons to measured irrep probabilities on the lattice. In Sec.\,\ref{sec:seamed-partition-function}, we derive the seamed partition functions for probability measurements in the representation basis. Sec.\,\ref{sec:mc_implementation} provides an overview of the algorithms used for numerical implementations of seamed Markov Chain Monte Carlo (MCMC) and measurements, which are then employed in Sec.\,\ref{sec:numerical_results} for measuring probabilities on a $4\times4\times4\times L_t$ lattice and extrapolating to the Hamiltonian limit $\xi \to \infty$. We conclude with an overview of results in Sec.\,\ref{sec:conclusions} and list possible future directions for study.

\section{Weak Coupling Perturbation Theory}\label{sec:weak-pt}

In this section, we evaluate  equal-time electric field squared correlators in both the continuum  and on a discretized spatial lattice, using weak-coupling ($g \to 0$) perturbation theory. Throughout this section, we will work in the temporal gauge $A_0=0$ with Gauss law $\nabla\cdot\mathbf{E} = 0$ imposed on the physical states.

\subsection{Continuum Perturbation Theory}\label{subsec:weak-pt-continuum}

The Yang-Mills Lagrangian in the continuum is given by
\begin{equation}
    \mathcal{L}_{\text{YM}} = - \frac{1}{4} F^a_{\mu \nu} F^{a \mu \nu},
\end{equation}
where the color field strength tensor is
\begin{equation}
    F^{a}_{\mu \nu} = \partial_\mu A_\nu^a - \partial_\nu A_\mu^a + g f^{abc} A_\mu^b A_\nu^c.
\end{equation}
As $g \to 0$, the Lagrangian can be approximated as a free color theory with gauge fields $A_i^a$ which fluctuate around the classical ground state $A_i^a(x) = 0$ up to a gauge transformation $A_i^a(x) \to A_i^a(x) + \partial_i\phi^a(x)$. Time independent gauge transformations are generated by the Gauss operator $Q^a(\mathbf{x}) = \nabla \cdot \mathbf{E}^a(\mathbf{x})$. In the free color limit, the Gauss operator generates the longitudinal modes of the gauge field. Therefore, the Gauss law constraint ensures that if transversality is imposed at initial time $t_0$, the gauge field will only have transversal modes at all later times.  Under this assumption of transversality, the mode expansion of the gauge field is given by
\begin{equation}
    A_i^a(x) = \int \frac{d^3p}{(2\pi)^3} \frac{1}{\sqrt{2p_0}} \sum_{\alpha=1,2} \epsilon_\alpha^i(\mathbf{p}) \left[ \hat{a}_\alpha(\mathbf{p},a) e^{-i p \cdot x} + \hat{a}_\alpha^\dagger (\mathbf{p},a) e^{i p \cdot x} \right],
\end{equation}
with the dispersion relation $p_0 = |\mathbf{p}|$ resulting from the equation of motion $(\partial_0^2 - \nabla^2) A_i^a = 0$. The $\hat a_\alpha,\hat a_\alpha^\dagger$ satisfy canonical commutation relations $[\hat{a}_\alpha(\mathbf{p},a), \hat{a}_\beta^\dagger(\mathbf{q},b)] = (2\pi)^3 \delta_{ab} \delta_{\alpha\beta} \delta^3(\mathbf{p}-\mathbf{q})$.

The electric field operator is related to the conjugate momentum as $E_i^a = - \Pi_i^a$; in the temporal gauge this is given by $E_i^a = - \partial_0 A_i^a$. The mode expansion of $E_i^a$ is therefore given by
\begin{equation}\label{eq:lattice-pt-2}
    E^a_i(x) = i \int \frac{d^3p}{(2\pi)^3} \sqrt{\frac{p_0}{2}} \sum_{\alpha=1,2} \epsilon_\alpha^i(\mathbf{p}) \left[ \hat{a}_\alpha(\mathbf{p},a) e^{-i p \cdot x} - \hat{a}_\alpha^\dagger (\mathbf{p},a) e^{i p \cdot x} \right].
\end{equation}
We are interested in evaluating the moments of the composite operator $E^2_i(x) \equiv \sum_{a=1}^{N^2-1} E^a_i(x) E^a_i(x)$. While the expectation value $\langle E^2_i(x) \rangle$ exhibits a quartic ultraviolet divergence $\sim \Lambda^4$, the connected two-point correlation function is finite and can be calculated up to leading order in $g$. The equal-time two-point electric color-contracted correlator is given by
\begin{equation}\label{eq:electric-color-correlator}
    \langle E_i^a(\mathbf{x}) E_j^a(\mathbf{y}) \rangle = \int \frac{d^3p}{(2\pi)^3} \frac{|\mathbf{p}|}{2} e^{i \mathbf{p} \cdot \mathbf{r}} \sum_{\alpha=1,2} \epsilon_\alpha^i(\mathbf{p}) \epsilon^j_\alpha(\mathbf{p}) \, ,
\end{equation}
where $\mathbf{r} = \mathbf{x} - \mathbf{y}$. Using the transverse projector relation $\sum_{\alpha=1,2} \epsilon_\alpha^i (\mathbf{p})\epsilon_\alpha^j(\mathbf{p}) = \delta_{ij} - \frac{p_i p_j}{\mathbf{p}^2}$, the tensor integral decomposes as
\begin{equation}
    \langle E_i^a(\mathbf{x}) E_j^a(\mathbf{y}) \rangle = \int \frac{d^3p}{(2\pi)^3} \frac{|\mathbf{p}|}{2} e^{i \mathbf{p} \cdot \mathbf{r}} \left(\delta_{ij} - \frac{p_i p_j}{\mathbf{p}^2} \right) = A(r) \delta_{ij} + B(r) \hat{r}_i \hat{r}_j,
\end{equation}
where $\hat{r}_i=r_i/r$. Contracting with $\delta_{ij}$ and $\hat{r}_i \hat{r}_j$ yields two scalar integrals $X_1 = 3A(r) + B(r)$ and $X_2 = A(r) + B(r)$. Evaluating these via spherical coordinates gives $X_1 = -1/(\pi^2 r^4)$ and $X_2 = 1/(\pi^2 r^4)$, and we find
\begin{equation}\label{eq:electric-color-correlator-continuum}
    \langle E_i^a(\mathbf{x}) E_j^a(\mathbf{y}) \rangle = -\frac{1}{\pi^2 r^4} \delta_{ij} + \frac{2}{\pi^2 r^4} \hat{r}_i \hat{r}_j.
\end{equation}
Using this, we can find the connected equal-time two-point electric energy correlator as
\begin{equation}
\begin{split}
    \langle E^2_i(\mathbf{x}) E^2_j(\mathbf{y}) \rangle_{\text{conn}} &= 2(N^2-1) \left[ \int \frac{d^3p}{(2\pi)^3} \frac{|\mathbf{p}|}{2} e^{i \mathbf{p} \cdot \mathbf{r}} \sum_{\alpha=1,2} \epsilon_\alpha^i(\mathbf{p}) \epsilon^j_\alpha(\mathbf{p}) \right]^2, \\
    &= \frac{2(N^2-1)}{\pi^4 r^8} \left( \delta_{ij} - 2\hat{r}_i \hat{r}_j \right)^2  ,
\end{split}
\end{equation}
where $N^2-1$ is the number of color generators (this is $8$ for $SU(3)$). Electric energy correlators therefore have UV divergence for small separations, indicating the presence of strong localized electric fluxes in the continuum theory. A representation basis description of the ground state would require all irreps to faithfully capture these small wavelength fluctuations in the continuum theory.

UV divergences at small length scales are naturally regularized in a lattice theory. Therefore, we only need finite number of irreps to capture local electric energy fluctuations on a lattice. Long wavelength observables such as few particle scattering between low-lying modes of the energy spectrum are expected to be minimally affected by this cut-off. To quantify the magnitude of electric energy fluctuations on the scale of lattice cut-off $r\sim a$, we will repeat similar perturbative calculations on the lattice.

\subsection{Lattice Discretization}\label{subsec:weak-pt-lattice}

Now we turn to the lattice regularized theory, working on a spatial lattice with continuous time (as is relevant for quantum simulations.)
With lattice spacing $a$,  integrals over spatial momenta are cut off at $\pm \pi/a$ on the edges of the Brillouin zone. In this section, we will calculate leading contributions to electric energy observables using lattice perturbation theory. 

\subsubsection{Operator Rescalings}
It is convenient to define rescaled lattice variables as $A_i^a{'} = ga A_i^a$ and $\Pi_i^a{'} = a^2\Pi_i^a/g$. In terms of these the Lagrangian is
\begin{equation}\label{eq:lattice-lagrangian}
    L = \frac{a}{2g^2}  \sum_{\mathbf{n},i} F^a_{0i}{'}F_{0i}^{a}{'} - \frac{a^3}{2g^2}  \sum_{\mathbf{n},i< j} \left(\Delta_i^+ A_j^{a}{'} - \Delta_j^+ A_i^{a}{'} \right)^2  ,
\end{equation}
and the canonical commutation relations are
\begin{equation}
    [\Pi_i^a{'}(a\mathbf{m}),A_i^a{'}(a\mathbf{n})] = - i \ \delta^3_{\mathbf{m},\mathbf{n}} .
\end{equation}
Here $\Delta^+_i$ is the forward finite difference operator defined as
\begin{equation}
\Delta^+_i f(\mathbf{x}) = \frac{f(\mathbf{x}+a\boldsymbol{i}) - f(\mathbf{x})}{a}.
\end{equation}
The gauge part of the Kogut-Susskind Hamiltonian can be found by a Legendre transform of the lattice Lagrangian in Eq.\,(\ref{eq:lattice-lagrangian}) and is given by
\begin{equation}
    H_{\text{KS}} = \frac{g^2}{2a} \sum_{\mathbf{x},i} E_i^a{'} (\mathbf{x})^2 + \frac{1}{g^2a} \sum_{\mathbf{x}, i<j} \text{Tr}\left[2I -W_{ij}{'} (\mathbf{x}) -W_{ij}{'}^\dagger(\mathbf{x}) \right]  ,
\end{equation}
where $W_{ij}{'}(\mathbf{x}) = U_{i}{'}(\mathbf{x}) U_j{'}(\mathbf{x}+ a\mathbf{i}) U_i{'}^\dagger (\mathbf{x} + a \mathbf{j}) U_j{'}^\dagger (\mathbf{x})$ with $U_i{'}(\mathbf{x}) = \exp\left(i A_i{'} (\mathbf{x}) \right)$ and the electric operator $E_i^a{'} = - \Pi_i^a{'}$. The relation between the rescaled and continuum electric operators is thereby given as
\begin{equation}\label{eq:lattice-E-scale}
    E_i^a{'} = - \frac{a^2}{g} \Pi_\mu^a =  \frac{a^2}{g} \ E_i^a .
\end{equation}
\subsubsection{Lattice Hamiltonian Perturbation Theory}
The magnetic plaquette term of the KS Hamiltonian can be expanded to quadratic order in $A_\mu^a{'}$. The linearized Hamiltonian splits into decoupled oscillators for each  color generator. Per color, the linearization is 
\begin{equation}
    H_{2} = \frac{g^2}{2a} \sum_{\mathbf{n},i} E_i^a{'}(\mathbf{n})^2 + \frac{1}{2a g^2} \sum_{\mathbf{n}, i<j} \left[ \Delta_i^+ A_j^a{'}(\mathbf{n}) - \Delta_j^+ A_i^a{'}(\mathbf{n}) \right]^2.
\end{equation}
In momentum space,  $\Delta_i^+$ acts as multiplication by $K_i(\mathbf{\tilde{p}})$ given by
\begin{equation}
    K_i(\mathbf{\tilde{p}}) = \frac{e^{i\tilde{p}_i} - 1}{a} = i \hat{p}_i e^{i\tilde{p}_i/2}  ,
\end{equation}
with dimensionless momentum $\tilde{p}_i$ and $\hat{p}_i a \equiv 2 \sin(\tilde{p}_i / 2)$. Similarly, $\Delta_i^-$ acts in momentum space as $-K_i^*(\mathbf{\tilde{p}}) = i \hat{p}_i e^{-i\tilde{p}_i/2}$. Using these  relations, we can express the magnetic term of $H_2$ in the momentum basis as (neglecting boundary terms)
\begin{equation}\label{eq:lattice-pt-1}
    \frac{1}{2} \left[\Delta_i^+ A_j^a{'} - \Delta_j^+ A_i^a{'} \right]^2 (\tilde{\mathbf{p}}) = \tilde{A}_j^a{'}(\mathbf{\tilde{p}}) \left[\hat{p}^2 \delta_{ij} - \hat{p}_i \hat{p}_j e^{i (\tilde{p}_i - \tilde{p}_j)/2} \right] \tilde{A}_i^a{'}(\mathbf{\tilde{p}})  ,
\end{equation}
where $\tilde{A}_i^a{'} (\mathbf{\tilde{p}})$ is the Fourier transform of the  gauge field $A_i^a{'} (\mathbf{n})$. Therefore, the magnetic term in Eq.\,(\ref{eq:lattice-pt-1}) factorizes via the diagonal unitary matrix $\Phi_{ij} = \delta_{ij} e^{i \tilde{p}_i / 2}$ into
\begin{equation}
    M = \Phi M_0 \Phi^\dagger, \quad \text{with} \quad M_0 = \hat{p}^2 \mathbb{I} - \hat{p}\hat{p}^T.
\end{equation}
Given orthogonal vectors $\{\mathbf{e}^1, \mathbf{e}^2\}$ in the plane perpendicular to the direction $\hat{\mathbf{p}}$, the physical transverse eigenvectors of $M_0$ are  $\epsilon^\lambda_i = \Phi_{ij} e_j^\lambda$ where $\lambda \in \{1,2\}$. The resulting completeness relation over physical transverse modes contains a phase factor:
\begin{equation}
    P_{ij}(\tilde{\mathbf{p}}) \equiv \sum_{\lambda=1,2} \epsilon_i^\lambda(\mathbf{p}) \epsilon_j^{\lambda *}(\mathbf{p})  = \delta_{ij} - \frac{\hat{p}_i \hat{p}_j}{\hat{p}^2} e^{i(\tilde{p}_i - \tilde{p}_j)/2}.
\end{equation}

The lattice dispersion relation follows from the discretized equation of motion $(\partial_0^2 - \Delta_i^- \Delta_i^+) A_\mu^a = 0$. The lattice dispersion relation is thereby given by
\begin{equation}
    p_0 =  |\hat{\mathbf{p}}| = \frac{2}{a}\sqrt{\sum_{k=1}^3 \sin^2(\tilde{p}_k/2)}  .
\end{equation}

The  discretized transverse projector and dispersion relations can now be used to find the expectation value of $E_i^2(x)$, which is now finite because it is UV-regulated by the lattice cutoff. The mode expansion in Eq.\,(\ref{eq:lattice-pt-2}),  restricted to the Brillouin zone, gives the following expression for $\langle E_i^2 (x) \rangle$:
\begin{equation}
    \langle E^2_i \rangle = (N^2-1) \int_{-\pi/a}^{\pi/a} \frac{d^3p}{(2\pi)^3} \ \frac{p_0}{2}  \left(1- \frac{\hat{p}_i^2}{\hat{p}^2} \right) \approx \frac{0.796 \times (N^2-1)}{a^4}  .
\end{equation}
Using the scaling relation (\ref{eq:lattice-E-scale}), we obtain
\begin{equation}\label{eq:lattice-pt-electric-energy-result}
    \langle E_i^2{'} \rangle \approx \frac{0.796 \times (N^2 -1 )}{g^2}  .
\end{equation}
The per link electric energy of the ground state, which is a proxy for small-wavelength electric fluctuations on the lattice, is finite as we expected. A suitable truncation of irrep electric energy $C_2(R)$ imposed on the link Hilbert space should therefore be able to reproduce short wavelength behavior of the lattice vacuum for finite $g$. The electric energy flux diverges as $g\to0$, thereby requiring higher and higher irreps as the continuum limit is approached.

The connected equal-time two-point electric energy correlator can also be similarly calculated as 
\begin{equation}
    \langle E^2_i(\mathbf{x}) E^2_j(\mathbf{y}) \rangle_{\text{conn}} = 2(N^2-1) \left[ \int_{-\pi/a}^{\pi/a} \frac{d^3p}{(2\pi)^3} \frac{p_0}{2} e^{i \mathbf{p} \cdot \mathbf{r}} \left( \delta_{ij} - \frac{\hat{p}_i \hat{p}_j}{\hat{p}^2} e^{i(p_i - p_j)a/2} \right) \right]^2.
\end{equation}
where the integral inside the square bracket is the electric color-contracted correlator
\begin{equation}
    \langle E_i^a(\mathbf{x}) E_j^a(\mathbf{y}) \rangle \equiv G_{ij}(\mathbf{r})  = \int_{-\pi/a}^{\pi/a} \frac{d^3p}{(2\pi)^3} \frac{p_0}{2} e^{i \mathbf{p} \cdot \mathbf{r}} \left( \delta_{ij} - \frac{\hat{p}_i \hat{p}_j}{\hat{p}^2} e^{i(p_i - p_j)a/2} \right)  .
\end{equation}
Let us calculate this correlator for three cases (refer to App.\,\ref{app:bessel} for conversions of the three-dimensional momentum integrals to one-dimensional Bessel integrals):
\begin{enumerate}[leftmargin=*]
    \item $\mathbf{r} = na \ \hat{e}_1$ and $i=j=1$:
    In this case, the integral simplifies to
    \begin{equation}\label{eq:g11-e1}
        G_{11}(n \hat{e}_1) = \frac{1}{a^4} \int_{-\pi}^{\pi} \frac{d^3p}{(2\pi)^3}  \ e^{inp_1} \ \left(\frac{\hat{p}_2^2 + \hat{p}_3^2}{2\hat{p}} \right)  .
    \end{equation}
    This integral can be written in terms of Bessel functions as 
    \begin{equation}
        G_{11}(n \hat{e}_1) = \frac{2}{a^4\sqrt{\pi}} \int_0^\infty \frac{dt}{\sqrt{t}} \ e^{-6t} \cdot I_n(2t) I_0(2t) \left[I_0(2t) - I_1(2t) \right]  .
    \end{equation}
    This is the collinear electric correlator along the lattice direction. This can be compared to the continuum electric correlator in Eq.\,(\ref{eq:electric-color-correlator-continuum}) at $r=na$ and $i=j$.
    \item $\mathbf{r} = na \ (\hat{e}_1+\hat{e}_2)$ and $i=j=1$:
    The integral in this case is given by
    \begin{equation}\label{eq:g11-e12}
        G_{11}(na \hat{e}_1 + na \hat{e}_2) = \frac{1}{a^4} \int_{-\pi}^{\pi} \frac{d^3p}{(2\pi)^3} \ e^{in (p_1+p_2)} \ \left(\frac{\hat{p}_2^2 + \hat{p}_3^2}{2\hat{p}} \right)  .
    \end{equation}
    In terms of Bessel functions, this function simplifies as
    \begin{multline}\label{eq:g12-e12}
        G_{11}(na \hat{e}_1 + na \hat{e}_2) = \frac{1}{2a^4\sqrt{\pi}} \int_0^{\infty} \frac{dt}{\sqrt{t}} \ e^{-6t} \ \left[4 I_n^2(2t)I_0(2t) - I_0(2t)I_n(2t)I_{n+1}(2t) \right. \\ \left. - I_0(2t)I_{n-1}(2t)I_n(2t) - 2I_n^2(2t) I_1(2t) \right]  .
    \end{multline}
    \item $\mathbf{r} = na \ (\hat{e}_1+\hat{e}_2)$ and $i=1$, $j=2$:
    The integral in this case is given by
    \begin{equation}
        G_{12}(na \hat{e}_1 + na \hat{e}_2) = -\frac{1}{a^4} \int_{-\pi}^{\pi} \frac{d^3p}{(2\pi)^3} \ e^{in (p_1+p_2)} \ \left(\frac{\hat{p}_1 \hat{p}_2}{2\hat{p}} e^{i(p_1-p_2)/2} \right)  .
    \end{equation}
    Again, in terms of Bessel functions, this is given as
    \begin{equation}
        G_{12}(na \hat{e}_1 + na\hat{e}_2) = \frac{1}{2a^4\sqrt{\pi}} \int_0^\infty \frac{dt}{\sqrt{t}} \ e^{-6t} \ I_0(2t) \left[I_{n+1}(2t) - I_n(2t) \right] \left[I_n(2t) - I_{n-1}(2t) \right] .
    \end{equation}
\end{enumerate}
The electric operator $E_i^a$ is linear in the limit of free color theory, so the electric operator in the $\hat{e}_1 + \hat{e}_2$ direction is given by $E_1^a + E_2^a$. One can infer then
\begin{equation}
    \langle E_{45^\circ}^a(\mathbf{x}) E_{45^\circ}^a(\mathbf{y}) \rangle = \langle (E_1^a(\mathbf{x}) + E_2^a(\mathbf{x})) (E_1^a(\mathbf{y}) + E_2^a(\mathbf{y}))\rangle = 2 \, G_{11}(\mathbf{r}) + 2 \, G_{12}(\mathbf{r})  .
\end{equation}
Matching this rotated collinear operator to its continuum counterpart in Eq.\,(\ref{eq:electric-color-correlator}) with $i=j$ and $r=\sqrt{2}na$ provides a test for rotational symmetry restoration. On the lattice, rotational invariance is explicitly broken at the UV scale -- the discretization distinguishes on-axis from diagonal directions, and the correlator deviates from its isotropic continuum form at short distances. 

\begin{figure}[ht!]
    \centering
    \includegraphics[width=0.8\linewidth]{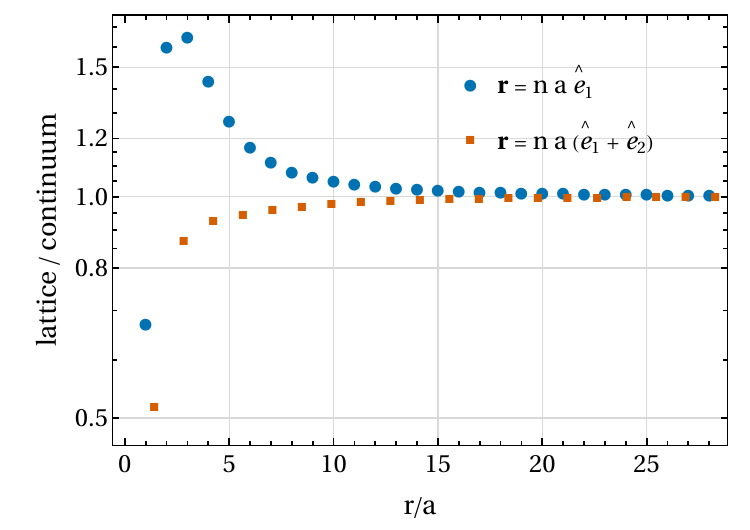}
    \caption{Ratio of two point collinear correlator $\langle E^2_i(\mathbf{x}) E^2_i(\mathbf{x+\mathbf{r}})\rangle$ computed in lattice and continuum PT. The blue circles denote the correlator calculated along the lattice link direction, and the yellow squares denote the correlator calculated $45^\circ$ to the link direction in the 1-2 plane.}
    \label{fig:weak-pt-vs-continuum}
\end{figure}

Fig.\,(\ref{fig:weak-pt-vs-continuum}) compares the ratio of the collinear electric correlator calculated along the lattice direction and $45^\circ$ to the lattice direction in the 1-2 plane with continuum up to leading order in $g$. Both of the lattice correlators collapse to the asymptotic scaling of $1/r^4$ of the continuum theory to within 3-5\% at $r=10 \,a$.

 The agreement between the lattice electric correlator and its continuum counterpart at the $\sim$ percent level for $r \gtrsim 10\,a$ holds even though the lattice calculation retains only a finite number of irreps in the representation basis, whereas the continuum theory requires the full tower of irreducible representations to resolve fluctuations at arbitrarily short scales. This tells us that higher irreps, which encode UV-sensitive, short-distance physics, are not needed to faithfully reproduce long-wavelength electric observables once $r$ is sufficiently large in lattice units. This is of course a statement of universality applied to a particular basis of truncation. Phenomena governed by electric flux at the scale $r$ such as string formation and string breaking can therefore be studied on a lattice with finite spacing $a \ll r$ without taking the continuum limit $a \to 0$.

This brings us to the question of how many irreps must be retained in practice. Perturbation theory can estimate the magnitude of the local electric energy stored by irreps, but it does not give access to the probability with which individual irreps are occupied in the link Hilbert space -- information that is essential for determining the optimal truncation order for a quantum simulation. In the following section, we develop a Monte Carlo method to compute these quantities directly, calculating the relative probabilities of irreps on individual links and of site-singlet configurations in a non-perturbative setup, and thereby establishing a quantitative criterion for where the representation basis can be safely truncated at a given lattice spacing.

\newcommand{\Cas}{C_2}

\section{Partition Function for Probabilities in the Representation Basis} \label{sec:seamed-partition-function}

Perturbation theory provides an estimate of the strength of local electric fluxes in the weak-coupling limit; however, quantum simulations of phenomenological interest do not operate in the asymptotically free weak-coupling regime -- they target physical volumes and lattice spacings where $g^2$ is of order unity and higher-order corrections are not suppressed. At such couplings, the perturbative expansion is unreliable, and the question of which irreps carry significant probability in the ground state must be answered non-perturbatively. Monte Carlo sampling of the Euclidean path integral provides exactly this access -- it sums contributions to all orders in $g^2$ and is valid across the full range of couplings relevant to near-term quantum hardware.

The central challenge is that standard Euclidean lattice Monte Carlo is formulated in the group element basis, which corresponds in Hamiltonian language to the magnetic basis: configurations are labeled by group elements $U \in SU(3)$ on each link, and the Boltzmann weight $e^{-S[U]}$ is diagonal in this basis. Electric observables, by contrast, are diagonal in the representation basis and off-diagonal in the magnetic basis -- they count the irrep content of a link state, which requires projecting a generic group element onto $SU(3)$ irreps via a character decomposition. Measuring such observables within a standard Monte Carlo simulation is therefore non-trivial. In this section we develop the theory for doing so, showing how a seam construction on measured links converts expectation values of irrep probabilities into character measurements that can be estimated by Monte Carlo sampling.

In order to achieve the Hamiltonian limit, anisotropy must be introduced in the lattice with the anisotropy factor $\xi = a_s/a_t$, where $a_s$ and $a_t$ are the lattice spacing in the spatial and temporal directions. The Hamiltonian limit is achieved as $\xi \to \infty$. To this effect, the following treatment will consider anisotropic lattices with dimensions $L \times L \times L \times L_t$ with periodic boundary conditions (PBC) to conserve translational invariance. 

\subsection{Transfer Matrix Review}

Let's work in temporal gauge (temporal link variables set to 1) and normalize the Haar measure so that  $\int_G dU=|G|=1$. Spatial links are denoted by $\ell$. Spatial plaquettes are denoted $P$ and $U_P$ is the plaquette operator associated with spatial plaquette $P$. Temporal plaquette operators $V_\ell(t)=U_\ell(t)\,U_\ell(t{+}1)^\dagger$ are associated with every spatial link $\ell$. Let there be $L_t$ timesteps with periodic boundary conditions such that $U_\ell(t{+}L_t)=U_\ell(t)$. $D^R$ denote representation matrices with $d_R=\dim R$, and the character function is $\chi_R(U)=\Tr D^R(U)$. 

The anisotropic Wilson action for $SU(3)$ is given by 
  \begin{align}
    S&=\frac{\beta}{3}\Big[(a_s^3a_t)\sum_{P}\frac{1}{a_s^4}\big(N-\Retr U_P\big)
      \;+\;(a_s^3a_t)\sum_{\ell,t}\frac{1}{a_s^2a_t^2}\big(N-\Retr V_\ell(t)\big)\Big]\nonumber\\
      &=\frac{\beta}{3}\Big[\frac1{\xi}\sum_{P}\big(N-\Retr U_P\big)
      \;+\;\xi\sum_{\ell,t}\!\!\big(N-\Retr V_\ell(t)\big)\Big]\nonumber,\\
  \end{align}
where $\xi=a_s/a_t$, and $\beta = 6/g^2$. The Hamiltonian limit is achieved by making the time direction continuous, i.e., $\xi\rightarrow\infty$. 

The partition function of the Wilson action can be expressed in a path integral form as
\begin{equation}\label{eq:path-integral-def}
  Z_{\text{W}}=\int\prod_{\ell,t}dU_\ell(t)\ \prod_{t=0}^{L_t-1}\Big[\prod_\ell K_{\text{W}}\big(V_\ell(t)\big)\Big]\,
  e^{-S_{\rm sp}[\{U(t)\}]}.
\end{equation}
where $S_{\text{sp}}[\{U(t) \}] = \sum_P S_P[\{U(t)\}]$ denotes the spatial Wilson action on the spatial slice with link values $\{U(t)\}$ and 
\begin{equation}
    S_P[\{U(t)\}] = \frac{\beta}{3 \xi} \big(N - \Retr U_P \big).
\end{equation}
$K_{\text{W}}(V_\ell(t))$ denotes the temporal part of the Wilson transfer operator such that
\begin{equation}
    K_{\text{W}}\big(V_\ell(t)\big) = 
\exp \left( -\frac{\beta \xi}{3} \big(N- \Retr V_\ell(t) \big) \right).
\end{equation}
It is beneficial in addition to study a modified action which converges to the KS Hamiltonian faster than the Wilson action in the limit $\xi \to \infty$ -- this is the heat kernel action~\cite{Menotti1981288}. The heat kernel action is constructed by modifying the temporal part of the Wilson action to match the electric operator of the KS Hamiltonian. The heat kernel path integral is given by
\begin{equation}
    Z_{\text{HK}} = \int\prod_{\ell,t}dU_\ell(t)\ \prod_{t=0}^{L_t-1}\Big[\prod_\ell K_{\text{HK}}\big(V_\ell(t)\big)\Big]\,
  e^{-S_{\rm sp}[\{U(t)\}]},
\end{equation}
where the temporal action is given by
\begin{equation}\label{eq:heat-kernel-action}
    K_{\text{HK}}\big(V_\ell(t)\big) = \sum_R d_R \chi_R(V_\ell (t))e^{-g^2C_2(R)/2\xi} .
\end{equation}
In the discussions in the rest of this section and subsequent sections, we will refer to $Z$ as the general path integral given by
\begin{equation}
    Z = \int\prod_{\ell,t}dU_\ell(t)\ \prod_{t=0}^{L_t-1}\Big[\prod_\ell K\big(V_\ell(t)\big)\Big]\,
  e^{-S_{\rm sp}[\{U(t)\}]},
\end{equation}
where the temporal part $K$ can either be $K_{\rm W}$ or $K_{\rm HK}$.

The connection of the Euclidean action to the Hamiltonian formalism is given by the transfer matrix. The Hilbert space of states is defined on a spatial slice $\mathcal H=\bigotimes_\ell L^2(G)$, with magnetic basis completeness relation
$\mathbf 1=\int\prod_\ell dU_\ell \,|\{U\}\rangle\langle\{U\}|$ for links $\ell$ that belong to a spatial slice. Then the transfer matrix is constructed by defining operators $\hat B$ and $\hat K$ such that
\begin{align}
  \hat B\,|\{U\}\rangle&\equiv e^{-S_{\rm sp}[\{U\}]}|\{U\}\rangle\nonumber\\
  \langle\{U\}|\hat K|\{U'\}\rangle&\equiv \prod_\ell K\big(U_\ell^\dagger U_\ell'\big),
  \end{align}
  so that the transfer matrix can be defined as
  \begin{align}\label{eq:transfer-matrix-def}
  \hat T\equiv\hat B^{1/2}\hat K\hat B^{1/2}
  \end{align}
  with magnetic basis matrix elements given by
  \begin{align}
  \langle\{U\}|\hat T|\{U'\}\rangle
  =e^{-\frac12 S_{\rm sp}[\{U\}]}\Big[\prod_\ell K\big(U_\ell U_\ell'^\dagger\big)\Big]
   e^{-\frac12 S_{\rm sp}[\{U'\}]}.
   \end{align}
   Because periodicity implies 
$  \prod_{t=0}^{L_t-1}e^{-\frac12 S_{\rm sp}[\{U(t)\}]}\,e^{-\frac12 S_{\rm sp}[\{U(t{+}1)\}]}
  =\prod_{t=0}^{L_t-1}e^{-S_{\rm sp}[\{U(t)\}]}$,
   the path integral in Eq.\,(\ref{eq:path-integral-def}) can be expressed in terms of the transfer matrix such that
  \begin{align}
  Z=\text{tr}\big[\hat T^{L_t}\big],
\end{align}
where the trace ``$\text{tr}$" is over the Hilbert space of spatial links.

\subsection{Measurement of Link Irreps}

The ground state expectation values of Hilbert space operators can be written as a Euclidean path integral using the transfer matrix formalism. Since $\hat T$ is hermitian and positive-definite, its spectrum consists of positive eigenvalues $\lambda_n$ defined as
$\hat T|n\rangle=\lambda_n|n\rangle$, $\lambda_0>\lambda_1\ge\cdots\ge0$. Therefore, the expectation value of an operator $\hat{O}$ is given by
\begin{equation}
  \ev{\hat O}\equiv\frac{\text{tr}[\hat T^{L_t}\hat O]}{\text{tr}[\hat T^{L_t}]}
  =\frac{\sum_n\lambda_n^{L_t}\langle n|\hat O|n\rangle}{\sum_n\lambda_n^{L_t}}
  \ \xrightarrow{\ L_t\to\infty\ }\ \langle\Omega|\hat O|\Omega\rangle .
\end{equation}

The operator we are interested in is the single-link irrep projector given as
\begin{align}
  \hat P_R=\sum_{a,b=1}^{d_R}|R;ab\rangle\langle R;ab|,
\end{align}
where $R$ is an irrep of $SU(N)$. Using
\begin{align}\label{eq:group-element-irrep-inner-products}
  \langle U|R;ab\rangle=\sqrt{d_R}\,D^R_{ab}(U),\qquad
  \langle R;ab|R';a'b'\rangle=\delta_{RR'}\delta_{aa'}\delta_{bb'}
  \end{align}
  we obtain magnetic basis elements of $\hat{P}_R$:
  \begin{align}
  \langle U|\hat P_R|U'\rangle
 &=d_R\sum_{a,b}D^R_{ab}(U)\,\overline{D^R_{ab}(U')}\nonumber\\
  &=d_R\sum_{a,b}D^R_{ab}(U)\,D^R_{ba}(U'^\dagger)\nonumber\\
  &=d_R\,\chi_R\big(UU'^\dagger\big).
  \end{align}
  The expectation value of the projector operator $\hat{P}_R$ on the link $\ell_0$ on a spatial slice can thereby be written as  
  \begin{align}
  \langle\{U\}|\hat P_R|\{U'\}\rangle
  =d_R\,\chi_R\big(U_{\ell_0}U_{\ell_0}'^\dagger\big)\prod_{\ell\ne\ell_0}\delta\big(U_\ell,U_\ell'\big).
\end{align}
The expectation value of the projector can therefore be measured by inserting $\hat{P}_R$ into the transfer matrix on the spatial link $\ell_0$ at slice $t_0$. Because the projector is not diagonal in the magnetic basis, we will have to insert an additional link state at the site $(\ell_0,t_0)$, leading to a doubling of links at this site. We will call the doubled link states $\ket{U_-}$ and $\ket{U_+}$. After this insertion, the partition function is given by
\begin{subequations}
\begin{align}
  Z_R&=\int\Big[\prod_{(\ell,t)\ne(\ell_0,t_0)}\!\!\!dU_\ell(t)\Big]dU_-\,dU_+\ W_{\rm bulk}\,W_{\rm def}(R),\\
  W_{\rm bulk}&=\Big[\prod_{(\ell,t)\notin\mathcal B}K\big(V_\ell(t)\big)\Big]
   \Big[\prod_{t\ne t_0}e^{-S_{\rm sp}[\{U(t)\}]}\Big]
   \Big[\prod_{P\not\ni\ell_0}e^{-S_P[\{U(t_0)\}]}\Big],\\
  W_{\rm def}(R)&=K\big(U_{\ell_0}(t_0{-}1)U_-^\dagger\big)\,K\big(U_+U_{\ell_0}(t_0{+}1)^\dagger\big)
   \prod_{P\ni\ell_0}e^{-\frac12 S_P(U_-)}e^{-\frac12 S_P(U_+)}\;d_R\chi_R\big(U_-U_+^\dagger\big),
   \end{align}
\end{subequations}
   where in the last line $ S_P(W)\equiv S_P[\{U(t_0)\}]\big|_{U_{\ell_0}(t_0)=W}$, i.e., the two half-step magnetic evolutions involve $U_-$ on the target link and $U_+$ on the target link, respectively. Notice that the doubled links $U_-$ and $U_+$ are only connected to the lower and upper quadrants of the time slice respectively. This can be thought of as a seam in the lattice at the position $(\ell_0,t_0)$. The probability of the irrep $R$ can now be calculated as
\begin{align} \label{eq:path-integral-irrep-formula}
  p(R)=\frac{Z_R}{Z}=\langle \hat P_R \rangle
   \ \xrightarrow{\ L_t\to\infty\ }\ \langle\Omega|\hat P_R|\Omega\rangle.
\end{align}

\subsection{Measurement of Site-Singlets}
The physical Hilbert space of the Hamiltonian formulation of $SU(N)$ gauge theories is constrained by the Gauss law $\nabla \cdot \mathbf{E} = 0$. This constraint can be used to build a ``reduced electric basis"~\cite{Ciavarella2021,quantum-circuits-draper} by contracting the internal irrep indices on link variables in the standard Byrnes-Yamamoto basis~\cite{Byrnes2006} into a site-singlet degree of freedom. Although this condition reduces the size of the physical Hilbert space significantly, the number of site-singlets still becomes substantial at high truncation. It is possible to probe the relative importance of the various site-singlets that appear in the physical Hilbert space by measuring the expectation value of a corresponding projection operator, using the same methods developed for measuring irrep content in the previous sections.

The site singlet projector $\hat P(1,\vec R,\Gamma)$ is constructed to project out the singlet with outer multiplicity $\Gamma$ from the tensor product between the irreps $\vec{R}$ that meet at a vertex $\mathcal{V}$. In the representation basis, the site singlet state is defined as
\begin{align}
|1,\vec R, \vec a, \Gamma\rangle = \sum_{\vec b} C^{1\Gamma}_{\vec b }(\vec R)\left(\otimes_i |R_i\, a_i\, b_i\rangle\right),
\end{align}
where $C_{\vec{b}}^{1\Gamma}$ is the Clebsch-Gordan coefficient for the singlet, and the projector is defined as a sum over all the color degrees of freedom such that
\begin{align}
\hat P(1,\vec R, \Gamma) = \sum_{\vec a}|1,\vec R, \vec a, \Gamma\rangle\langle 1,\vec R, \vec a, \Gamma|.
\end{align}
The matrix elements of this projector in the group element basis can again be computed using the inner product relations in Eq.\,(\ref{eq:group-element-irrep-inner-products}) to obtain
\begin{align}\label{eq:site-singlet-formula}
\langle\{ U\}|\hat P(1,\vec R,  \Gamma)|\{U'\}\rangle &= \sum_{\vec a} \langle \{U\}| \sum_{\vec b} C^{1\Gamma}_{\vec b }(\vec R)\left(\otimes_i |R_i\, a_i\, b_i\rangle\right) \sum_{\vec b'} C^{1\Gamma}_{\vec b' }(\vec R)\left(\otimes_i \langle R_i\, a_i\, b'_i|\right)|\{U'\}\rangle\nonumber\\
&=\sum_{\vec b}   \sum_{\vec b'} C^{1\Gamma}_{\vec b }(\vec R)C^{1\Gamma}_{\vec b' }(\vec R)\left(\prod_{i\in \mathcal{V}} d_{R_i}\,D^{R_i}_{b_i b_i'}(U_iU^{\dagger '}_i)\right) \prod_{j \notin \mathcal{V}} \delta(U_j, U_j{'}).
\end{align}
Note that one must be careful about the direction of the link variables $U_j$ while using this formula --- a convention needs to be chosen to ensure that all of the link variables that meet at a vertex either point out or point into the vertex. We will choose the convention wherein all the link variables involved in the singlet point out of the vertex; any link variable $U_j$ that points into the vertex is replaced by its inverse $U_j^\dagger$ in Eq.\,(\ref{eq:site-singlet-formula}).  Additionally, we will use Clebsch-Gordan coefficients $C^{1\Gamma}$ which are real and orthonormal, $\sum_{\vec b}C^{1\Gamma}_{\vec b}C^{1\Gamma'}_{\vec b}=\delta^{\Gamma\Gamma'}$. Then the $\hat P(1,\vec R,\Gamma)$ are commuting orthogonal projectors, $\sum_{\vec R,\Gamma}\hat P(1,\vec R,\Gamma)=\hat G_x$ is the Gauss-law projector at $x$. 

The transfer matrix for the singlet projector now consists of $2d$ seams at the vertex $(x_0,t_0)$, where $x_0$ is a vertex on the spatial slice at time $t_0$. Let $\{U_-\}$ and $\{U_+\}$ be the sets of seams on which the site singlet projector is measured. Then, we have that
\begin{align}
\langle\{U_-\}|\hat B^{1/2}\,\hat P(1,\vec R,\Gamma)\,\hat B^{1/2}|\{U_+\}\rangle
=e^{-\frac12 S_{\rm sp}[\{U_-\}]}\,\langle\{U_-\}|\hat P(1,\vec R,\Gamma)|\{U_+\}\rangle\,
e^{-\frac12 S_{\rm sp}[\{U_+\}]},
\end{align}
Therefore, a plaquette containing $k\ge1$ doubled links therefore splits cleanly into an all-minus half and an all-plus half. 

Reading off the path integral form (seam bonds $\mathcal B\equiv\{(\ell_i,t_0{-}1),(\ell_i,t_0)\}_{i=1}^{N}$):
\begin{subequations} \label{eq:site-singlet-seam-integral}
\begin{align}
Z(1,\vec R,\Gamma)&=\int\Big[\prod_{(\ell,t)\ne(\ell_i,t_0)}\!\!\!dU_\ell(t)\Big]\prod_{i=1}^{N}dU_{i,-}\,dU_{i,+}\ W_{\rm bulk}\,W_{\rm def}(1,\vec R,\Gamma),\\
W_{\rm bulk}&=\Big[\prod_{(\ell,t)\notin\mathcal B}K\big(V_\ell(t)\big)\Big]
\Big[\prod_{t\ne t_0}e^{-S_{\rm sp}[\{U(t)\}]}\Big]
\Big[\prod_{P\,\cap\,{\mathcal{V}}=\emptyset}e^{-S_P[\{U(t_0)\}]}\Big],\\
W_{\rm def}(1,\vec R,\Gamma)&=\Big[\prod_{i=1}^{N}K\big(U_{\ell_i}(t_0{-}1)U_{i,-}^\dagger\big)\,K\big(U_{i,+}U_{\ell_i}(t_0{+}1)^\dagger\big)\Big]\nonumber\\
&\qquad\times\Big[\prod_{P\,\cap\,\mathcal{V}\ne\emptyset}e^{-\frac12 S_P(\{U_-\})}\,e^{-\frac12 S_P(\{U_+\})}\Big]\,
\mathcal K_\Gamma,\\
\mathcal K_\Gamma(\vec R;\{V_i\})&\equiv\sum_{\vec b,\vec b'}C^{1\Gamma}_{\vec b}(\vec{R})\,C^{1\Gamma}_{\vec b'}(\vec{R})\,\prod_{i\in \mathcal{V}}d_{R_i}\,D^{R_i}_{b_ib_i'}(V_i), \label{mathK}
\end{align}
\end{subequations}
with $V_i\equiv U_{i,-}U_{i,+}^\dagger$ and $S_P(\{U_\pm\})$ the slice-$t_0$ plaquette action evaluated on $\{U_\pm\}$. Then
\begin{align}\label{eq:path-integral-site-singlet-formula}
p(1,\vec R,\Gamma)=\frac{Z(1,\vec R,\Gamma)}{Z}
\ \xrightarrow{\ L_t\to\infty\ }\ \langle\Omega|\hat P(1,\vec R,\Gamma)|\Omega\rangle.
\end{align}

\subsection{Discretization Error}
\label{subsec:richardson_extrapolation}
Connection between the transfer matrix defined in Eq.\,(\ref{eq:transfer-matrix-def}) and the Hamiltonian is made by taking the limit $a_t \to 0$, i.e., by making the time direction continuous. The transfer matrix for the Wilson action flows towards the KS Hamiltonian in this limit~\cite{creutz-transfer-matrix}:
\begin{equation}
    \hat{T}_{\text{W}} = \mathcal{N} \exp \left(- a_t \hat{H}_{\text{KS}} + \mathcal{O}(a_t^2) \right),
\end{equation}
where $\mathcal{N}$ is an irrelevant constant. This expression can be compared to the effective lattice Hamiltonian $\hat{T}_{\text{W}} = \exp \left(- a_t H_{\text{W}} \right)$ to get that
\begin{equation}
    H_\text{W} = H_{\text{KS}} + \mathcal{O}(a_t).
\end{equation}
The convergence of the lattice Hamiltonian to the KS Hamiltonian can be improved by instead using an improved heat kernel action~\cite{Menotti1981288} which has the same electric term as the KS Hamiltonian, i.e.,
\begin{align}
  \hat{K}_{\text{HK}}=e^{-a_t H_E^{\rm KS}},  \qquad H_E^{\rm KS} = \frac{g^2}{2a_s}\hat C_2.
  \end{align}
Therefore, the only source of error between the heat kernel lattice Hamiltonian and the KS Hamiltonian is second-order trotter error given by
\begin{equation}
    \hat{T}_{HK} = \exp \left(-a_t H_{\text{HK}} \right) = e^{-\frac{a_t}2 H_B}e^{-a_t H_E}e^{-\frac{a_t}2 H_B} = \exp \left(-a_t H_{\text{KS}} +  \mathcal{O}(a_t^3) \right),
\end{equation}
so that
\begin{equation}
    H_{\text{HK}} = H_{\text{KS}} + \mathcal{O}(a_t^2).
\end{equation}
Accordingly, the corrections to the ground states of the Wilson and heat kernel actions are respectively given as
\begin{subequations}
    \begin{align}
        \ket{\Omega_{\text{W}}} &= \ket{\Omega_{\text{KS}}} + \mathcal{O}(a_t), \\
        \ket{\Omega_{\text{HK}}} &= \ket{\Omega_{\text{KS}}} + \mathcal{O}(a_t^2).
    \end{align}
\end{subequations}
Therefore, in the limit $a_t\to 0$, or equivalently $\xi \to \infty$, the probability estimators for the two actions converge as
\begin{subequations}
    \begin{align}
        &\bra{\Omega_{\text{W}}(\xi)} \hat{P}_R \ket{\Omega_\text{W}(\xi)} = p_{\text{W}} (R,\xi) = \bra{\Omega_{\text{KS}}} \hat{P}_R \ket{\Omega_{\text{KS}}} + \frac{\mathcal{E}_{\text{W}}^{(1)}(R)}{\xi} + \frac{\mathcal{E}_{\text{W}}^{(2)}(R)}{\xi^2}  + \mathcal{O}(1/\xi^3),\\
        &\bra{\Omega_{\text{HK}}(\xi)} \hat{P}_R \ket{\Omega_\text{HK}(\xi)} = p_{\text{HK}}(R,\xi) = \bra{\Omega_{\text{KS}}} \hat{P}_R \ket{\Omega_{\text{KS}}} + \frac{\mathcal{E}_{\text{HK}}^{(2)}(R)}{\xi^2} + \mathcal{O}(1/\xi^3).
    \end{align}
\end{subequations}

\section{Monte Carlo Implementation and Observables}
\label{sec:mc_implementation}

In this section, we will describe the algorithms used to produce Markov chain samples used for measurements in the representation basis using seamed actions described in Sec.\,\ref{sec:seamed-partition-function}. Furthermore, we will describe the thermalization and measurement protocols to achieve decorrelated data points from the otherwise highly correlated Markov chain samples.  
\subsection{Path Integral Sampling in the Representation Basis}
\label{subsec:pi_sampling}

To measure observables in the representation basis, the link variables involved in the measurement must be doubled as described in Sec.\,\ref{sec:seamed-partition-function}. Following the doubling of these target links, expectation values of representation-basis operators are evaluated by Monte Carlo sampling over the configuration weights using Markov chain processes with detailed balance to iteratively approach the target equilibrium distribution. 

A subtlety arises when evaluating the normalized expectation values. In Eqs.\,(\ref{eq:path-integral-irrep-formula}) and (\ref{eq:path-integral-site-singlet-formula}), the weights of the seamed partition function $Z_R$ must be divided by the full, unseamed partition function $Z$ to yield physical probabilities. However, direct Monte Carlo sampling only provides access to the seamed partition function weights normalized relative to the open seam partition function $Z_1$. Consequently, rather than directly computing $Z_R/Z$, the measured value is the ratio $Z_R/Z_1$, representing the relative probability of obtaining the irrep $R$ with respect to the trivial irrep $R=1$:
\begin{equation}\label{eq:mc_probability}
    p_\text{MC}(R) \equiv  \frac{Z_R}{Z_1} = \frac{p(R)}{p(1)}.
\end{equation}
To recover absolute probabilities we must divide the relative probabilities $p_{\text{MC}}(R)$ by their sum $\sum_{R'} p_{\text{MC}}(R')$, such that
\begin{equation}\label{eq:real-probability-using-mc}
    p(R) = \lim_{R'' \to \infty} \frac{p_{\text{MC}}(R)}{\sum_{R'=1}^{R''} p_{\text{MC}}(R')}.
\end{equation}
An important source of error in the extraction of absolute probabilities is the finite truncation of the irreps that are measured. Only probabilities above some noise threshold can be measured precisely. Because the error that is introduced by the truncation of the irrep $R'$ is proportional to $\chi_{R'} d_{R'}$, the error can be significant if the dimension of $R'$ is too large. 

\begin{figure}[ht!]
    \centering
    \includegraphics[width=\linewidth]{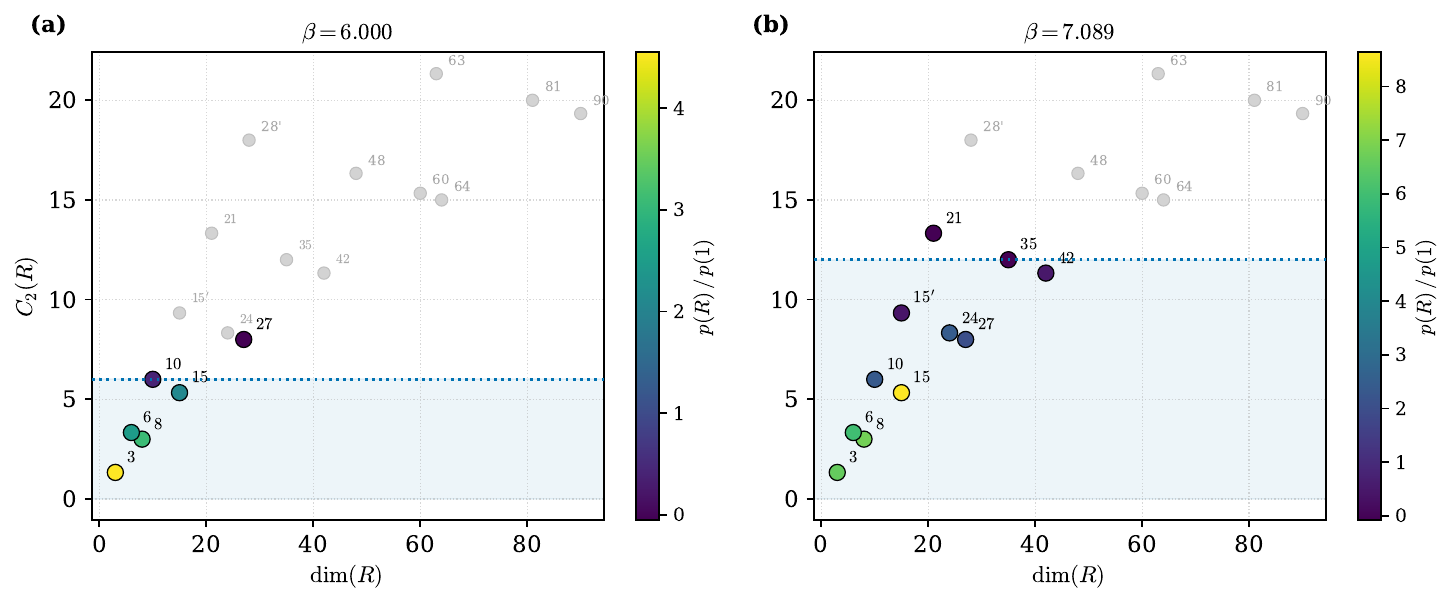}
    \caption{Active link irreps in Monte Carlo sampling of the seamed lattice at (a) $\beta=6.0$ $(g=1)$ and (b) $\beta=7.089$ $(g=0.92)$ for the heat kernel action at $\xi = 5.0$. The dots in the blue shaded region denote the irreps with character expectation values greater than $10^{-2}$ in every run of the MCMC. The characters with magnitude less than $10^{-2}$ cannot be measured reliably because the signal is overpowered by statistical noise.} 
    \label{fig:irrep-populations}
\end{figure}

Let us estimate the magnitude of error introduced by truncating the sum in the denominator of Eq.\,(\ref{eq:real-probability-using-mc}). (The details of the MC analysis and results are given in subsequent sections; here we just quote the relevant values of $p_{\text{MC}}(R)$ and use them for illustration.) For a given magnitude of statistical noise, only a few irreps with dominant population in the link Hilbert space can be measured, while the signals from suppressed irreps is indistinguishable from statistical noise. Fig.\,(\ref{fig:irrep-populations}) shows the various active irreps in Monte Carlo measurements of link probabilities at $\beta=6.0$ ($g=1$) and $\beta=7.089$ ($g=0.92$). While only irreps containing at most three index tensors are active for $\beta=6.0$, higher index tensors become more important for small lattice spacings. For $\beta = 6.0$, the largest irrep that can be measured with reasonable accuracy is the Dynkin label $(3,0) + \text{h.c.}$ with dimension $10$; the order of the character expectation value for the next higher irrep $(2,2)$ with $d_{R'} = 27$ can be conservatively estimated to be around the statistical noise $\sim 10^{-2}$. The value of $\sum_{R'} P_{\text{MC}} (R')$ for the heat kernel action is $\sim 13$ thereby making the systematic error in the probability measurement due to truncation around $27\times0.01/ 13 \approx 2\%$.  The truncation for $\beta \sim 7.0$ includes representations up to the Dynkin label $R = (5,0) + (0,5)$ with dimension $35$. The character expectation value for the next higher irrep with dimension $21$ is again estimated to be of order $\sim 10^{-2}$ and $\sum_{R'} P_{\text{MC}} (R') \sim 34 $, making the systematic error $\sim 0.6\%$. These systematic errors are therefore comparable to or less than the statistical noise in the sampling, which is around $2\%$.

The same normalization procedure described above for measurement of link irreps is followed to obtain normalized probabilities of site-singlets:
\begin{equation}
    p_{\text{MC}} (1,\Vec{R},\Gamma) = \frac{Z(1,\Vec{R},\Gamma)}{Z_1},
\end{equation}
and
\begin{equation}
    p(1,\Vec{R},\Gamma) = \frac{p_{\text{MC}}(1,\Vec{R},\Gamma)}{\sum_{(\vec{R}',\Gamma')}p_{\text{MC}}(1,\Vec{R}',\Gamma')}.
\end{equation}

\subsection{Link Update Algorithms}
\label{subsec:mc_algorithms}

There are many link update that can be used to get closer towards the target action, the easiest of which is the Metropolis algorithm. However, Metropolis updates converge  slowly towards the target distribution. Heat bath updates, for which proposed link update is always accepted, converge faster~\cite{creutz-heatbath}. For $SU(3)$, one chooses  $SU(2)$ subgroups of link matrices in the fundamental representation of $SU(3)$ and performs heat bath updates over each of the three $SU(2)$ subgroups~\cite{cabibbo-pseudo-heatbath}. 

This ``pseudo-heatbath" algorithm can  be applied easily to the Wilson action, because this action is expressed solely in the fundamental representation of $SU(3)$, so finding $SU(2)$ subgroups is straightforward. It becomes more non-trivial to evaluate the heat kernel action under such separation of subgroups because it involves the computation of characters for irreps higher than the fundamental representation of $SU(3)$. Therefore, for the heat kernel temporal action, we will employ a hybrid Metropolis algorithm where a link is first proposed using heat-bath sampling of the Wilson action followed by a Metropolis acceptance/rejection step using modified heat kernel weights. 

For each action and a given value of $\beta$ and $\xi$,  the lattice size chosen is $4\times 4 \times 4 \times L_t$, where $L_t = 24 \cdot \xi$. The temporal direction is changed with the anisotropy factor so as to keep the physical length of the temporal direction constant for different $\xi$. The seams are chosen to be in the time slice that lies in the middle of the two ends of the time direction. For measurement of the link irreps, 8 seams are chosen such that each of the seams is separated by at least two spatial coordinates; the characters are averaged over these seams in order to stabilize the fluctuations in the results. For measurement of site-singlets, a vertex is chosen in the middle of the lattice and all the links attached to the vertex are seamed.

At each step in the Markov chain process, a new link is proposed using either the Cabibbo-Marinari algorithm or a hybrid heatbath + Metropolis algorithm for the Wilson and heat kernel actions. The updated links are chosen from plaquettes in a checkerboard pattern to maintain detailed balance. Both of these algorithms equilibrate much faster than the ordinary Metropolis algorithm. We allow the system to thermalize for 20,000 steps, after which measurements are made every 10 steps for 100,000 total steps. 

Below, we describe in detail the link-update algorithms used for both the Wilson and heat kernel actions.

\begin{enumerate}[label=\Alph*.,leftmargin=*]
    \item \textbf{Wilson Action:} Since the Wilson action consists only of traces over the fundamental irrep, it is possible to use the Cabibbo-Marinari algorithm.
    \begin{enumerate}[label=\arabic*.,leftmargin=*]
        \item For the link variable $U$ that needs to be updated, the staple sum $A$ is computing by the computing the matrix
        \begin{equation}
           A = \sum_{\ell_1,\ell_2,\ell_3 \in P(U)} U_{\pm \ell_1} U_{\pm \ell_2} U_{\pm \ell_3} 
        \end{equation}
        where the sum is over all the plaquettes that contains $U$, and the orientation of the links $\pm \ell$ is decided relative to the orientation of $U$ in the plaquette operator. 

        \item Now, the aim is to draw a new link $U' = R U$ from a heat bath distribution of the form 
        \begin{equation}
            d P(U') = dU' \ \exp \left(\frac{\beta}{3} \Retr \ (RUA) \right)
        \end{equation}
        The collective product $U A$ acts as a heat bath for the unitary $R$. Next $2\times2$ subblocks of the $SU(3)$ matrices $R$ and the product $W = UA$ are selected, called $R_{\text{sub}}$ and $W_{\text{sub}}$. The term in the Wilson action $\Retr (RW)$ splits cleanly into terms that depend only on the multiplications of the sub-blocks without any interference from terms not in the sub-block. The problem now reduces to drawing $SU(2)$ samples from the pseudo-heatbath distribution
        \begin{equation}
            d P(R_{\text{sub}}) = d R_{\text{sub}} \exp \left( \frac{\beta \sqrt{\det W_{\text{sub}}}}{3} \Retr (R_{\text{sub}} V_{\text{sub}}) \right)
        \end{equation}
        where $V_\text{sub}$ is the $SU(2)$ matrix obtained by normalizing $W_{\text{sub}}$ such that
        \begin{equation}
            V_{\text{sub}} = \frac{W_{\text{sub}}}{\sqrt{\det W_\text{sub}}}.
        \end{equation}
        Conventional heatbath methods for $SU(2)$ can be used to draw $R_{\text{sub}}$ from this probability distribution~\cite{creutz-heatbath}. 

        \item The proposed link can be derived from $R_{\text{sub}}$ by embedding $R_{\text{sub}}$ into a $3\times3$ matrix $R$ such that $R$ acts as an identity matrix outside the subspace of the embedded block. The updated link is given by $U' = RU$, as defined before. 

        \item Repeat this for all three $SU(2)$ subgroups to obtain the update matrices $R$, $S$, and $T$. Then, the proposed link at the end of all three updates is $U' = (RST)  \cdot U$.
        
    \end{enumerate}
    \item \textbf{Heat Kernel Action:} The heat kernel action consists of characters higher than the fundamental trace; therefore, it becomes untenable to split it into subgroups of $SU(2)$. We therefore use a hybrid Metropolis update algorithm described as follows:
    \begin{enumerate}[label=\arabic*.,leftmargin=*]
        \item Use the Cabibbo-Marinari algorithm to find a heat bath candidate for all the links (using the Wilson action).
        \item The hybrid acceptance probability is given by
        \begin{equation}
            T_A(U'|U) = \min \left(1, \frac{T_0(U|U') \exp(-S[U'])}{T_0(U'|U) \exp(-S[U])} \right)
        \end{equation}
        where $T_0(U'|U)$ is the a priori selection probability of $U'$ given the initial configuration $U$. In our case, this apriori selection probability is the Wilson path integral weight given by
        \begin{equation}
            T_0(U'|U) = \exp \left(-S_W[U'] \right),
        \end{equation}
        so we obtain
        \begin{equation}
            T_A(U' | U) = \min \left(1, \frac{\exp (-S_W[U]) \exp(-S_{\text{KS}}[U'])}{\exp(-S_W[U']) \exp(-S_{\text{KS}}[U])} \right).
        \end{equation}
        Because the spatial part of the Wilson and heat kernel actions are the same, they cancel out in the above expression such that
    \begin{equation}
        T_A(U'|U) = \min \left(1, \frac{K_{\text{HK}}[U']/K_W[U']}{K_{\text{HK}}[U]/K_W[U]} \right).
    \end{equation}
    The weights used in the Metropolis-Hastings acceptance/rejection step is therefore given by the ratio of the heat kernel and Wilson actions. The acceptance rate of this Hybrid algorithm is found to be above $90\%$, which is much higher than the ordinary acceptance rate of $50\%$ when starting with a randomly generated guess. 
    \end{enumerate}
    
\end{enumerate}

\subsection{Thermalization and Measurement Protocol}
\label{subsec:measurement_protocol}

A central requirement of MCMC algorithm is that the chain must have reached its stationary distribution, controlled by the target action, before any measurements are recorded. Starting from an arbitrary initial configuration, the chain undergoes a transient relaxation during which the distribution of configurations is still influenced by the choice of initial state and has not yet converged to equilibrium. After thermalization is achieved, successive configurations along a Markov chain are still not statistically independent. Since each configuration is generated by a local update of the previous one, nearby configurations are correlated over a characteristic timescale known as the integrated autocorrelation time $\tau_{\text{int}}$. Successive measurements should be spaced such that the interval between each measurement is larger than the typical autocorrelation time of the observables of interest. 

 \begin{figure}[h!]
        \centering
        \begin{subfigure}{0.48\textwidth}
            \includegraphics[width=\linewidth]{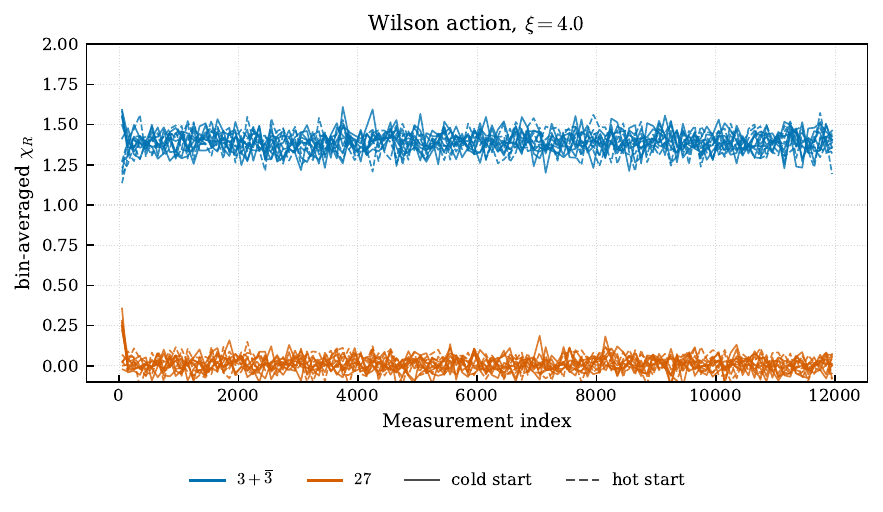}
            \caption{}
        \end{subfigure}
        \begin{subfigure}{0.48\textwidth}
            \includegraphics[width=\linewidth]{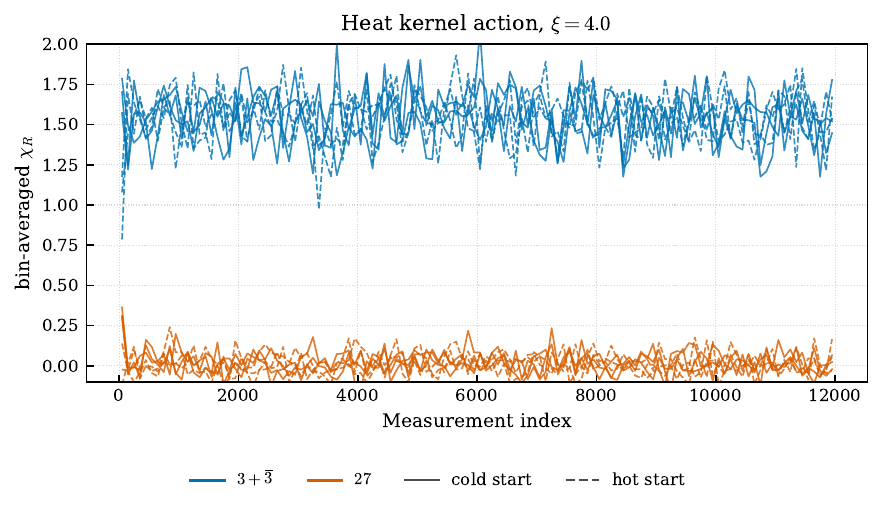}
            \caption{}
        \end{subfigure}
        \begin{subfigure}{0.48\textwidth}
            \includegraphics[width=\linewidth]{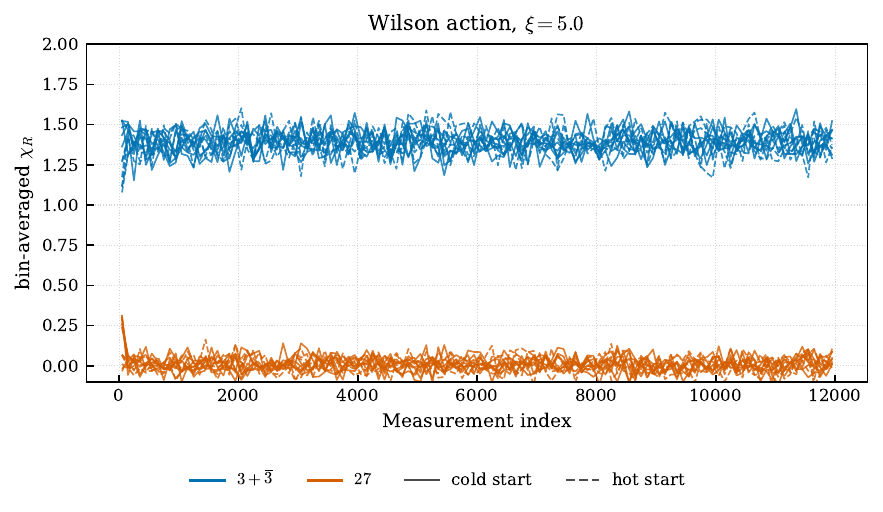}
            \caption{}
        \end{subfigure}
        \begin{subfigure}{0.48\textwidth}
            \includegraphics[width=\linewidth]{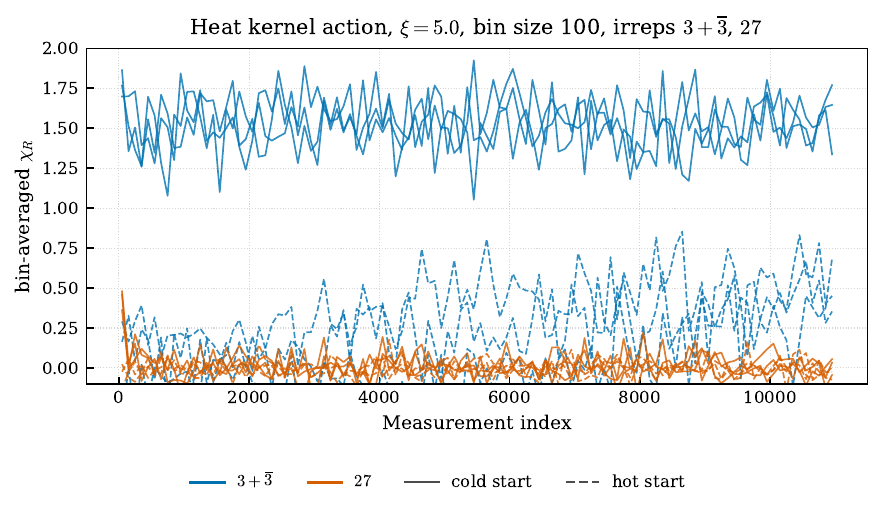}
            \caption{}
        \end{subfigure}
        \caption{Each panel plots the binned averages with bin size of 100 as a function of measurement index for irreps $3+\bar{3}$ and $27$ for the following cases ($\beta = 6.0$ in all cases) -- (a) Wilson action with $\xi=4.0$; (b) heat kernel action with $\xi=4.0$; (c) Wilson action with $\xi=5.0$; and (d) heat kernel action with $\xi=5.0$.}
        \label{fig:thermalization}
    \end{figure}

Here, we lay out the thermalization and measurement protocol employed to combat such effects:
\begin{enumerate}[leftmargin=*]
   
    \item \textbf{Initialization}: To assure convergence of measurements independent of initial conditions and correlated updates, we repeat the measurements over multiple Markov chains initialized in one of two configurations: an ordered (cold) start, in which all link variables are set to the identity matrix $U_\mu(x) = 1$; or a disordered (hot) start, in which each link is drawn independently and uniformly from the gauge group. For the Wilson action, we run 12 independent chains -- 6 hot and 6 cold -- while for the more computationally intensive heat kernel action, we run 6 chains -- 3 hot and 3 cold.
    \begin{figure}[ht!]
        \centering
        \includegraphics[width=\linewidth]{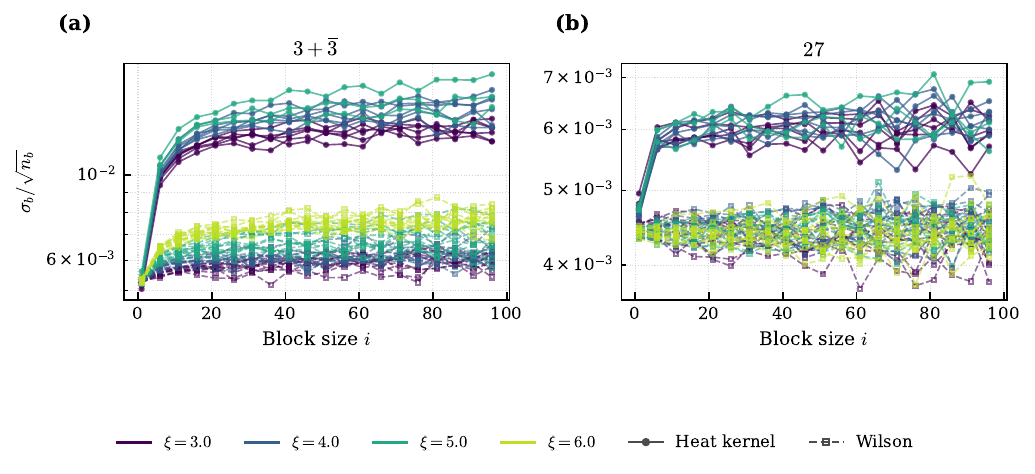}
        \caption{Jackknife errors plotted as a function of bin size plotted for $\xi \geq 3.0$ with $\beta=6.0$ for the Wilson and heat kernel actions for irreps $3+\bar{3}$ and $27$.}
        \label{fig:block-convergence}
    \end{figure}

    \item \textbf{Thermalization}: The system should be allowed to equilibrate for a long enough time for all the chains to converge towards a common distribution. Fig.\,(\ref{fig:thermalization}) plots local binned averages over a bin size of 100 as a function of measurement index (measurements are taken every 10 steps) for the $3+\bar{3}$ and $27$ irreps. The local averages equilibrate quickly for both the hot and cold chains except when $\xi=5.0$ for the heat kernel action -- here the hot chains get stuck and evaluate irrep expectation values close to 0. In our analysis, we discard the first 20,000 steps (i.e., the first 2000 measurements) to reduce transient thermalization effects; additionally, we discard the hot chains for the heat kernel action with $\xi=5.0$.

    \begin{figure}[ht!]
        \centering
        \begin{subfigure}{0.48\linewidth}
            \includegraphics[width=\linewidth]{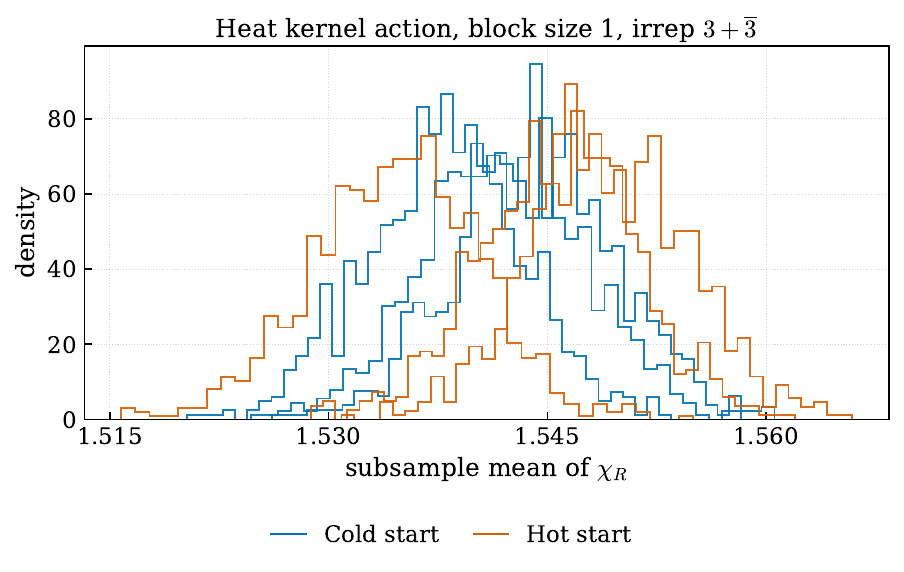}
            \caption{}
            \label{fig:mean-distribution-ks-3-bin1}
        \end{subfigure}
        \begin{subfigure}{0.48\linewidth}
            \includegraphics[width=\linewidth]{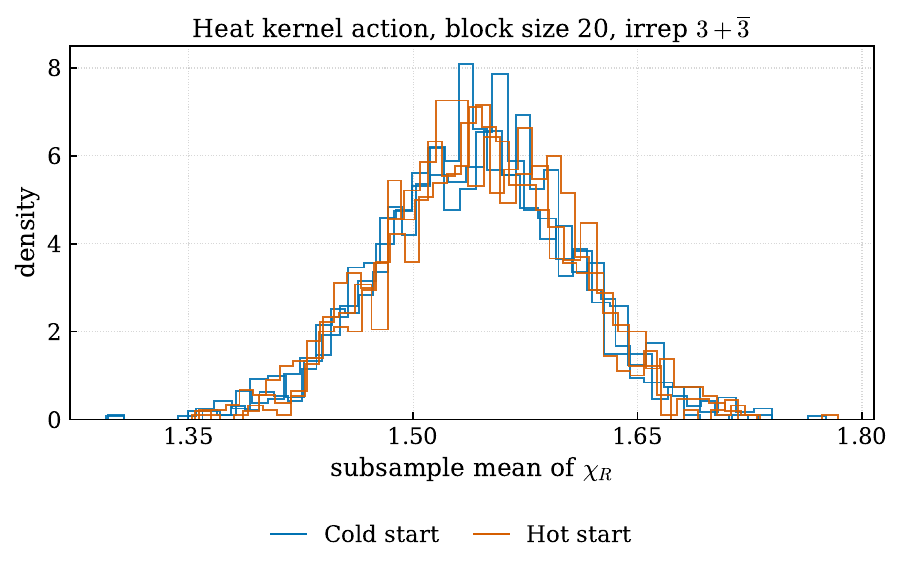}
            \caption{}
            \label{fig:mean-distribution-ks-3-bin20}
        \end{subfigure}
        \begin{subfigure}{0.48\linewidth}
            \includegraphics[width=\linewidth]{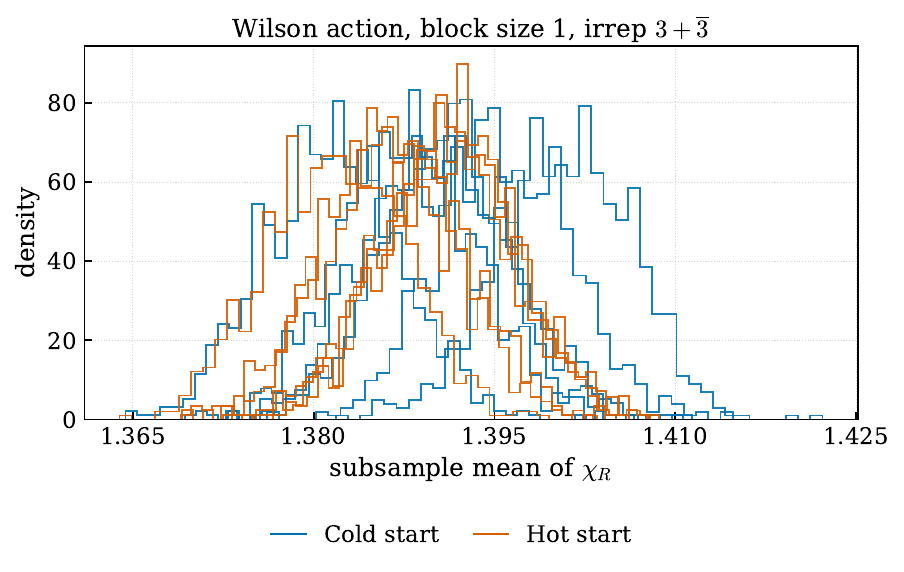}
            \caption{}
            \label{fig:mean-distribution-wilson-3-bin1}
        \end{subfigure}
        \begin{subfigure}{0.48\linewidth}
            \includegraphics[width=\linewidth]{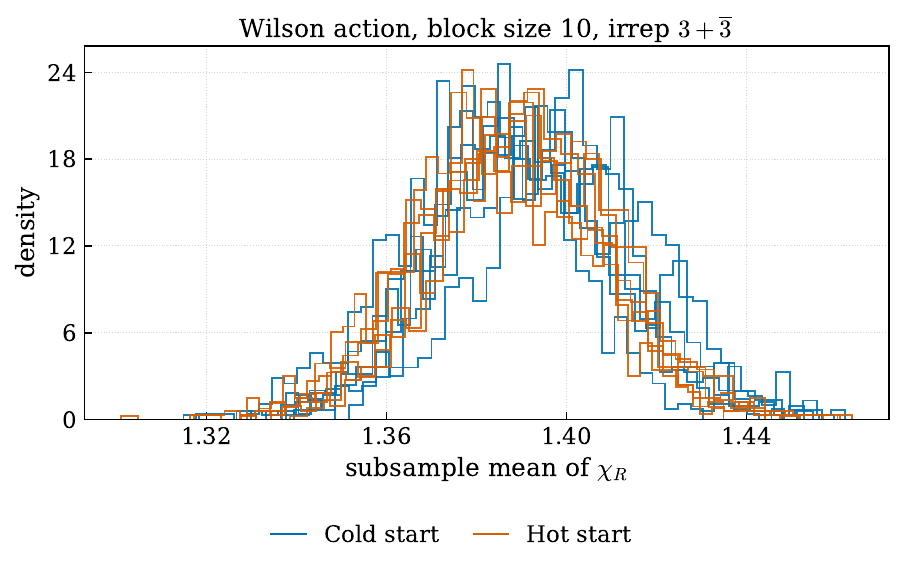}
            \caption{}
            \label{fig:mean-distribution-wilson-3-bin10}
        \end{subfigure}
        \caption{Mean distributions plotted for different chains of the Wilson and heat kernel actions with $\xi=4.0$ and $\beta=6.0$ for irreps $3+\bar{3}$ using bootstrap sampling wherein $N$ random samples are drawn with replacement 1000 times ($N$ is the number of elements in the original sample space) and the mean is computed for each draw. Panels (a) and (c) plot the mean distributions of the heat kernel and Wilson actions without binning, and panels (b) and (d) plot the mean distributions of said actions after binning in bin sizes of 20 and 10.}
        \label{fig:mean-distributions}
    \end{figure}

    \item \textbf{Measurement}: Measurements for both actions are recorded every 10 steps for 100,000 steps after the initial discard of pre-thermal measurements, yielding 10,000 post-thermal measurements. Only 90,000 post-thermalization steps are done for the case of the heat kernel action when $\xi=5.0$ because of high computing times. To ascertain decorrelation amongst samples, we bin each chain into uniformly-sized blocks and plot the standard deviation of irrep probabilities obtained by jackknife sampling as a function of bin size, as shown in Fig.\,(\ref{fig:block-convergence}). The standard deviation ($\sigma$) starts out at around $5\times 10^{-3}$ for bin size 1 and subsequently asymptotes to a larger value as the bin size is increased, indicating that bins of this size are decorrelated. For the Wilson action, $\sigma$ asymptotes for bin sizes of $\approx 10$, while for the heat kernel action it asymptotes for bin sizes $\approx 20$. Calculating expectation values without binning can lead to underestimation of the standard deviation. This effect can be seen by plotting the mean distributions computed by bootstrap sampling; this is plotted for $\xi=4.0$ and $\beta=6.0$ for the $3+\bar{3}$ irrep in Fig.\,(\ref{fig:mean-distributions}). Figs.\,(\ref{fig:mean-distribution-ks-3-bin1}) and (\ref{fig:mean-distribution-wilson-3-bin1}) plot the mean distributions across chains without any binning -- it is apparent from these figures that the variation of central values of means across chains is comparable or larger than the width $\sigma$ of the distributions, indicating that the error in means is underestimated. When, instead, the mean distributions are plotted after binning, the distributions converge towards a central distribution with a broader width, the central value and width of which can be computed without ambiguity.
\end{enumerate}

We employ the following strategy to compute expectation values based on the above discussion:
\begin{enumerate}
    \item The first 2000 ``pre-thermalization" measurements from all chains are discarded.
    \item Subsequent measurements are binned using bin size 10 for Wilson action and bin size 20 for heat kernel action.
    \item Binned samples from all chains are pooled together and means and standard deviations are computed on this collective sample space using jackknife sampling.
\end{enumerate}

\subsection{Irrep Character Evaluation and $(p,q)$ Matrix Construction}
\label{subsec:irrep_construction}
The formula for the expectation value of link irrep $R$ in Eq.\,(\ref{eq:path-integral-irrep-formula}) consists of evaluating the character $\chi_R$. Furthermore, fast evaluations of the characters of high irreps are required to efficiently evaluate the heat kernel action in Eq.\,(\ref{eq:heat-kernel-action}). We use the Weyl formula to evaluate characters~\cite{fulton1991representation} which consists of evaluating eigenvalues in the fundamental representation. Since $U \in SU(3)$ is unitary with $\det U = 1$, its characteristic polynomial is fully determined by $\operatorname{Tr}U$ and its conjugate, which reduces eigenvalue extraction to a single depressed cubic solvable in closed form via Cardano's formula. The implementation is further sped up in C by memoizing powers of the resulting eigenvalues in a look-up table which are reused across all evaluated irreps $R$.

The formula for the expectation value of the site-singlet in Eq.\,(\ref{eq:site-singlet-seam-integral}) requires the evaluation of the matrix representation $D^{(p,q)}(V)$. This is done by using the fundamental matrices $D^{(1,0)}(V) = V$ and anti-fundamental matrices $D^{(0,1)}(V) = \overline{V}$ that can be directly obtained from the MC implementation. 

To obtain the representation matrix in irrep $(p,q)$, we take the Kronecker product $V^{\otimes p}\otimes\overline{V}^{\otimes q}$ which can be done efficiently in Python. This tensor product is in the representation $3^p \otimes (3^*)^q$ which contains the irrep $(p,q)$ in its direct sum decomposition. This component of the direct sum can be projected out by pre-computing projection matrices $Q$ that are independent of $V$. The following are the steps for computing the matrix $Q$:
\begin{enumerate}[leftmargin=*]
    \item Use the Gell-Mann matrices in the fundamental representation $T^a_{(1,0)}$ to build the lie algebra basis in the $3^p\otimes (3^*)^q$ representation such that
    \begin{equation}
    T^a_{3^p\otimes 3^{*q}} = \sum_{i=1}^p\left(1^{\otimes(i-1)}\otimes T^a_{(1,0)}\otimes 1^{\otimes(p+q-i)}\right) + \sum_{j=1}^q\left(1^{\otimes(p+j-1)}\otimes  T^a_{(0,1)}\otimes 1^{\otimes(q-j)}\right),
    \end{equation}
    where $T^a_{(0,1)} = -(T^a_{(1,0)})^*$. The Casimir operator in this representation is then given by
    \begin{equation}
    C_2 = \sum_a \left(T^a_{3^p \otimes 3^{*q}}\right)^2.
    \end{equation}

    \item The Casimir operator can be diagonalized; there will be  exactly $\frac{1}{2}(p+1)(q+1)(p+q+2)$ eigenvectors that have  eigenvalue $\frac{1}{3}(p^2+q^2+3p+3q+pq)$. These eigenvectors form the basis of the $(p,q)$ irrep subspace. We can use these eigenvectors to project the Gell-Mann matrices onto the $(p,q)$ subspace to obtain the matrices $T^a_{p,q}$.

    \item We need to ensure that the irrep matrix $D^{(p,q)}(V)$ is in the same basis as the basis used for calculating the Clebsch-Gordan coefficients in Eq.\,(\ref{eq:site-singlet-seam-integral}). In our implementation of building the site-singlet operator, we compute Clebsch-Gordan coefficients using \texttt{pyclebsch}~\cite{B-truncation}. This codebase uses the Gelfand-Tsetlin (GT) basis to construct the CGCs. To find the GT basis, we can simultaneously diagonalize (see App.\,\ref{app:gt-basis}):  
    \begin{enumerate}
        \item The isospin operator $(T^1_{p,q})^2 + (T^2_{p,q})^2 + (T^3_{p,q})^2$ ;
        \item The first Cartan generator $J_z^{(1)} = T^3_{p,q}$; and
        \item The second Cartan generator $J_z^{(2)} = -T^3_{p,q}/2 + \sqrt{3}T^8_{p,q}/2$
    \end{enumerate}
    The eigenvectors of this simultaneous diagonalization procedure are now in the GT-basis and will form the projection operator $Q$ after appropriately fixing the signs of the eigenvectors in step 4.

    \item Additionally, care must be taken to match the Condon-Shortley-type phases used by \texttt{pyclebsch}~\cite{quantum-circuits-draper}. In order to find this phase, we find the GT pattern for each anti-fundamental slot for each eigenvector found in step 3. We then add up the Condon-Shortley phase obtained from the \texttt{pyclebsch} code for the GT pattern of each anti-fundamental slot. This cumulative phase of all the anti-fundamental slots is the correct phase of the eigenvector.

    \item Once the eigenvectors of the $(p,q)$ irrep subspace are found with the correct quantum numbers and phases, $Q$ can be constructed such that each column of $Q$ corresponds to an eigenvector found in step 4.
    
\end{enumerate}
Using $Q$, the $D^{(p,q)}(V)$ can be projected onto as
\begin{equation}
    D^{(p,q)}(V) = Q^\dagger V^p\otimes \overline{V}^q Q .
\end{equation}

\section{Numerical Results and Continuum Scaling}
\label{sec:numerical_results}
In this section, we will present results for link irrep and site-singlet probabilities measured on 4 dimensional anisotropic lattices of size $4\times4\times4\times L_t$, where $L_t$ is varied with the anisotropy factor $\xi$ as $L_t = 24 \cdot \xi$ in order to keep the physical length of the time direction constant across varying anisotropies. The Hamiltonian limit is recovered by extrapolating the data to $\xi \to \infty$. The Monte-Carlo algorithms used to generate samples and the measurement protocols used across the samples are described in Sec.\,(\ref{sec:mc_implementation}).

\subsection{Link Irrep and Site-Singlet Measurements}\label{subsec:link-irrep-measurement}
\begin{figure}[ht!]
    \centering
    \begin{subfigure}{0.48\linewidth}
        \includegraphics[width=\linewidth]{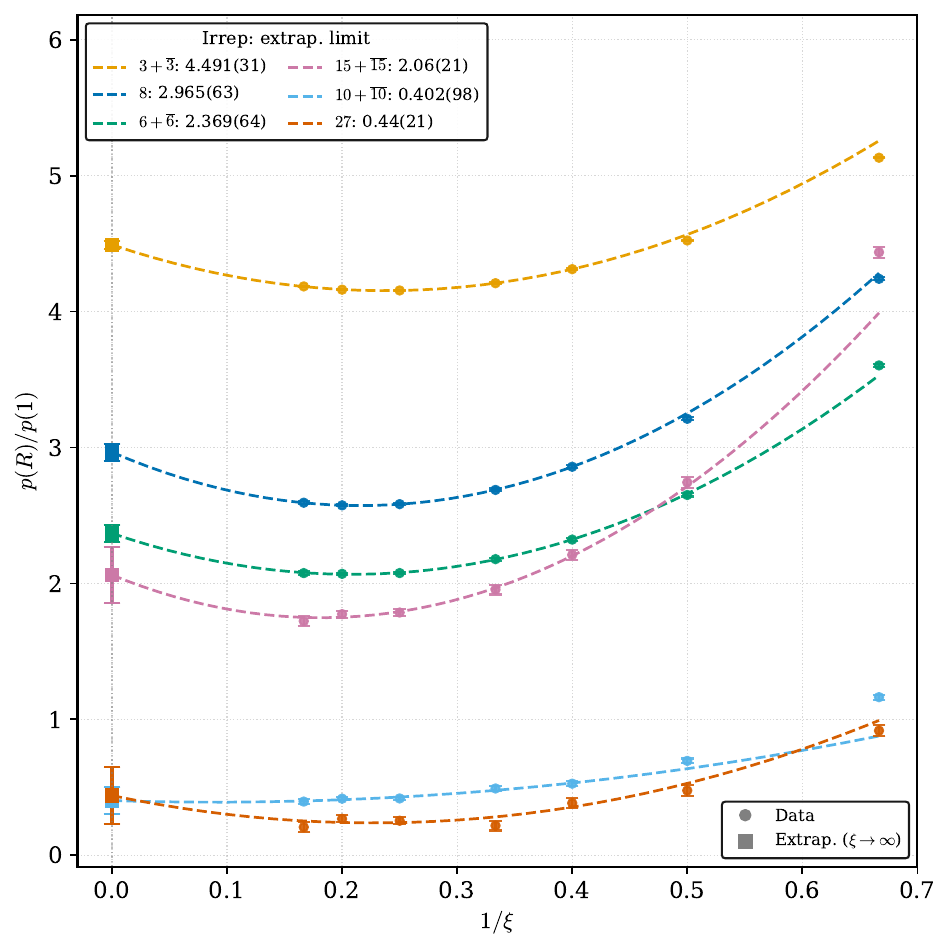}
        \caption{}
    \end{subfigure}
    \begin{subfigure}{0.48\linewidth}
        \includegraphics[width=\linewidth]{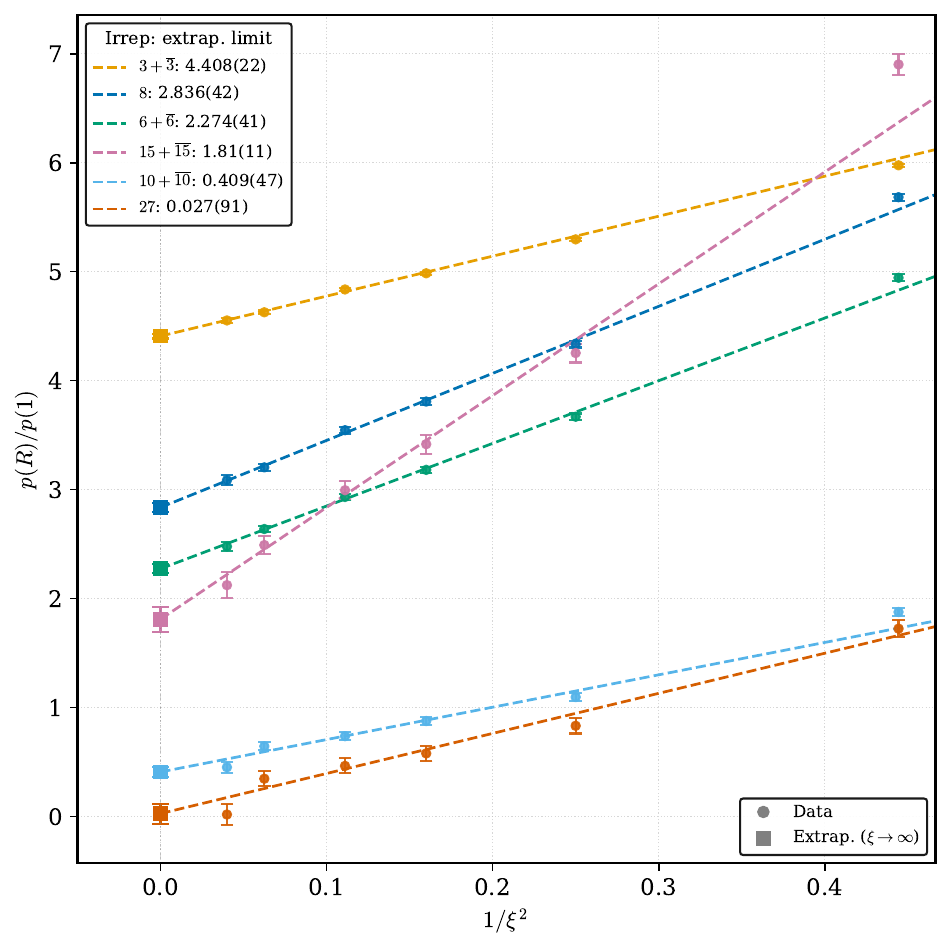}
        \caption{}
    \end{subfigure}
    \caption{Relative link irrep probability measurements for (a) Wilson and (b) heat kernel actions at $\beta=6.0$ plotted for varying anisotropies $\xi$ and extrapolated to the Hamiltonian limit $\xi\to\infty$ using (a) quadratic extrapolation with the curve $a+b/\xi+c/\xi^2$, and (b) linear extrapolation with the curve $a + b/\xi^2$.}
    \label{fig:link-irrep-xi-extrapolation}
\end{figure}
The link irrep probabilities $p(R)/p(1)$ were evaluated for the Wilson action for anisotropies $\xi = \{1.0,1.5,2.0,2.5,3.0,4.0,5.0,6.0\}$ at $\beta =6.0$ and for the heat kernel action for $\xi=\{1.0,1.5,2.0,2.5,3.0,4.0,5.0\}$; the $\xi=6.0$ point was omitted for the heat kernel action because of the high computational cost. As discussed in Sec.\,(\ref{subsec:richardson_extrapolation}), the ground state of the Wilson action converges to the KS Hamiltonian limit as $a + b/\xi + c/\xi^2$ while the heat kernel action converges faster with the trend $a + b/\xi^2$. We use the method of least-squares to fit the respective curves using data points $\xi > 2.0$ and report the extrapolated values and the $\chi^2/\rm dof$ of all the link irreps calculated in Table.\,\ref{tab:extrapolated_irreps}.  The p-value for the difference in extrapolated probabilities using the two actions was calculated using Welch's t-test~\cite{NIST:EngineeringStatisticsHandbook:3.1} -- all values evaluate above the conventional threshold p-value of 5\%, indicating that the two actions agree within these uncertainties.

\begin{table}[htbp]
    \centering
    \caption{Extrapolated values of link irrep probabilities for the Wilson and heat kernel actions, with the $\chi^2/\mathrm{dof}$ of each extrapolation ($\mathrm{dof}=2$) and a two-tailed Welch's $t$-test comparing the two actions (degrees of freedom $\nu$ from the Welch--Satterthwaite formula).}
    \label{tab:extrapolated_irreps}
    \begin{tabular}{c cc cc ccc}
        \toprule
        & \multicolumn{2}{c}{Wilson} & \multicolumn{2}{c}{Heat kernel} & \multicolumn{3}{c}{Welch's t-test} \\
        \cmidrule(lr){2-3} \cmidrule(lr){4-5} \cmidrule(lr){6-8}
        Irrep & Extrapolated Limit & $\chi^2/\mathrm{dof}$ & Extrapolated Limit & $\chi^2/\mathrm{dof}$ & $t$ & $\nu$ & $p$-value \\
        \midrule
        $\mathbf{3}$  & $4.49135 \pm 0.03116$ & $0.087$ & $4.40849 \pm 0.02224$ & $1.292$ & $2.164$  & $3.62$ & $0.104$ \\
        $\mathbf{8}$  & $2.96519 \pm 0.06288$ & $0.013$ & $2.83609 \pm 0.04184$ & $0.579$ & $1.709$  & $3.48$ & $0.173$ \\
        $\mathbf{6}$  & $2.36850 \pm 0.06424$ & $0.230$ & $2.27381 \pm 0.04095$ & $0.497$ & $1.243$  & $3.40$ & $0.293$ \\
        $\mathbf{15}$ & $2.06183 \pm 0.20733$ & $0.817$ & $1.81028 \pm 0.11289$ & $0.662$ & $1.066$  & $3.09$ & $0.363$ \\
        $\mathbf{10}$ & $0.40179 \pm 0.09777$ & $0.995$ & $0.40900 \pm 0.04657$ & $2.213$ & $-0.067$ & $2.86$ & $0.951$ \\
        $\mathbf{27}$ & $0.43906 \pm 0.20849$ & $3.364$ & $0.02703 \pm 0.09146$ & $2.408$ & $1.810$  & $2.74$ & $0.177$ \\
        \bottomrule
    \end{tabular}
\end{table}

It is apparent that the extrapolations extracted from the Wilson and heat kernel samples have competing sources of error:
\begin{enumerate}
    \item The Wilson action has light computational overhead and the samples generated have small autocorrelation times. Therefore, it is easy to iterate several hot+cold MC chains for this action and obtain decorrelated measurements with very low statistical noise. However, Wilson observables have slow convergence to the Hamiltonian limit which introduces significant error in the $\xi \to \infty $ extrapolation.

    \item The heat kernel action has a much higher computational cost and the samples generated have significantly larger autocorrelation times. Therefore, fewer samples are generated and the statistical noise is high. However, convergence of heat kernel observables to the Hamiltonian limit is much faster and linear extrapolation to $\xi \to \infty$ adds very little error. 
\end{enumerate}
 Despite differences in measured data and extrapolation techniques, we nevertheless reach the same conclusion from both actions -- irreps which consist of tensors with 4 or more indices are highly suppressed in the link Hilbert space for $\beta=6.0$. Moreover, similar analyses of link irrep population for hyperfine lattices with $\beta\sim7.0$ (see Fig.\,\ref{fig:irrep-populations}) indicate that three-index tensors such as the 15 dimensional representation with Dynkin labels $R=(2,1) + (1,2)$ become increasingly important as the continuum limit is approached. Therefore, a threshold truncation that faithfully reproduces the UV structure of the KS ground state must absolutely contain the two-index tensors and likely some contributions from the 3-index tensors, if the Kogut-Susskind Hamiltonian is used.

 \begin{figure}[ht!]
    \centering
    \includegraphics[width=\linewidth]{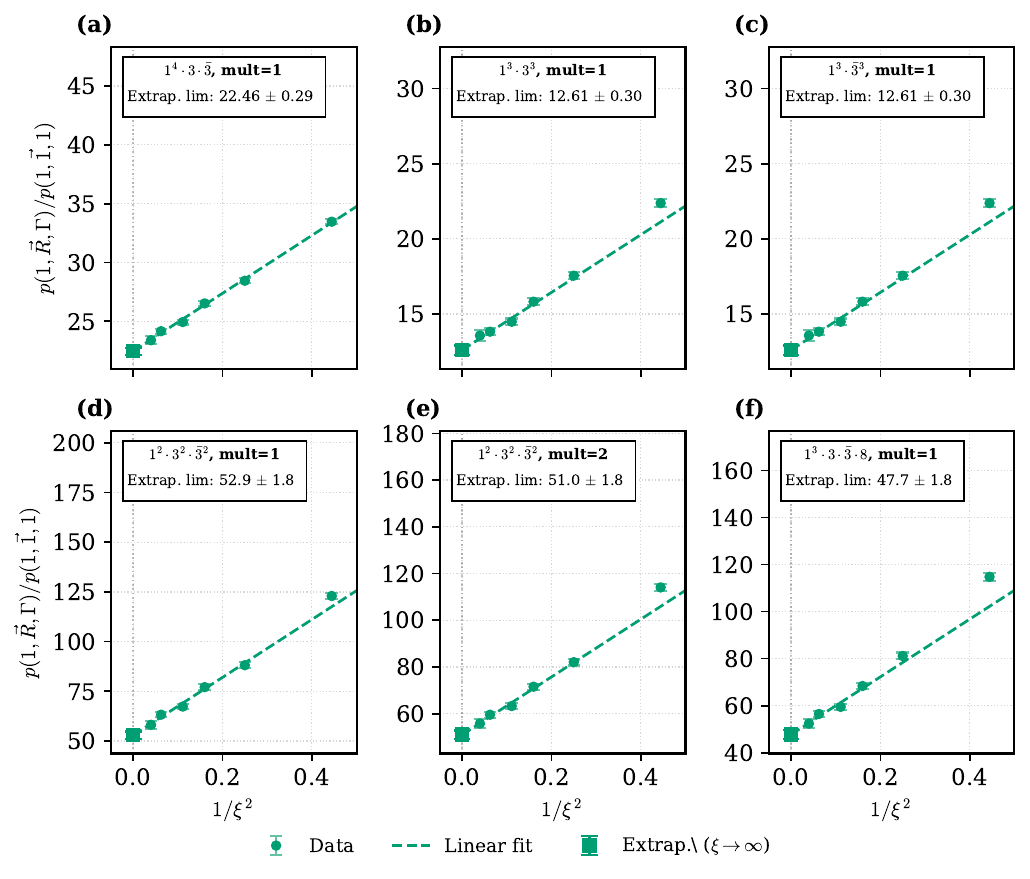}
    \caption{Measurement of relative probabilities of different configuration of site-singlets for the heat kernel action for varying anisotropies $\xi$ and their extrapolation to $\xi \to \infty$ using the linear extrapolation model $a+b/\xi^2$. Note here that the probability $p(1,\vec{R},\Gamma)$ is measured relative to the probability of the site singlet formed by trivial irreps $p(1,\vec{1},1)$.} 
    \label{fig:singlet-probabilities}
\end{figure}

The occurrence of an irrep in the link wavefunction is highly influenced by neighboring links which share a common vertex. This is because of the constraint of Gauss law for physical states in the Hilbert space, which requires the tensor product of all the irreps that meet at a vertex to form a $SU(3)$ singlet. Site-singlets can be categorized by the total electric energy contained in the constituent irreps. The average electric energy density is expected to be finite at finite lattice spacing, motivating a truncation of this total electric energy $B$ allowed at a vertex; we refer to a bound on this vertex energy as ``$B$"~\cite{B-truncation}.  $B$-truncation schemes provide much finer control over the truncated KS Hilbert space dimension  and plaquette matrix elements. 

Measurement of the abundance of site-singlets in the KS ground state can inform which $B$-truncation (or other related prunings of site singlets) are needed to faithfully reproduce the KS ground state and low-lying glueball states for few-particle scattering. In addition, measurements of the relative probabilities of singlets that have the same value of $B$ can be used to further refine the $B$-truncation scheme by cutting out site-singlets that have suppressed abundance. For example, the tensor product $8\otimes8\otimes8\otimes8\otimes8\otimes8$ has 145 singlets, but it is possible that some or perhaps even most of these singlets are unimportant, and therefore can be truncated from the Hilbert space without significant loss of accuracy. 

Vertex singlet probabilities present a greater measurement challenge than single-link irrep probabilities. The total probability is distributed across a combinatorially large number of singlet configurations, so each individual contribution is small and, in most cases, falls below the statistical noise floor of $1/\sqrt{N_{\text{meas}}} \sim 0.02$ at our sample size of $N_{\text{meas}} = 3000$ (10,000 measurements over 6 chains binned in blocks of 20 gives 3000 independent measurements). The signal is further suppressed by the dimension factor $\prod_R d_R$ appearing in Eq.\,(\ref{eq:site-singlet-seam-integral}), which weights each singlet by the product of the dimensions of its constituent irreps; singlets built from higher-dimensional representations are therefore doubly suppressed -- both because of their dimensional pre-factor and high value of $B$. As a result, only the few dominant low-$B$ singlets produce a signal distinguishable from statistical noise at our current statistics. Resolving contributions from singlets with larger $B$ would require a substantially larger Monte Carlo ensemble.

Fig.\,(\ref{fig:singlet-probabilities}) shows the extrapolated probabilities of the 6 dominant vertex configurations that were measured. The vertex with a $3$ and $\overline{3}$ and the rest in trivial irreps with $B=2.67$ has the most dominant population (Fig.\,(\ref{fig:singlet-probabilities}a)). Singlets formed by combinations of fundamental and anti-fundamental irreps are significantly more abundant than the trivial singlet, indicating the presence of strong fundamental flux loops in the KS vacuum. Panels (b) and (c) of Fig.\,(\ref{fig:singlet-probabilities}) measure the abundance of singlets formed by combinations of triplets $3\times3\times3$ and $\bar{3}\times\bar{3}\times\bar{3}$, and these probabilities are equal, as expected because of charge-conjugation symmetry of the KS Hamiltonian. The tensor product $3\otimes3\otimes\bar{3}\otimes\bar{3}$ has two singlets which are measured in panels (d) and (e) of Fig.\,(\ref{fig:singlet-probabilities}). The two multiplicities, labeled mult=1 and mult=2, are computed in a mutually orthogonal and Young-symmetrized basis~\cite{B-truncation}; the first multiplicity contains combinations that are symmetric under exchange of identical tensors while the second multiplicity contains anti-symmetric combinations. The extrapolated probabilities indicate that both multiplicities appear in the ground state with similar probabilities.

\subsection{Comparison to the Perturbation Theory and Tadpole Corrections} \label{subsec:pt_mc_comparison}
We can compare the values of the electric energy obtained from lattice perturbation theory in Eq.\,(\ref{eq:lattice-pt-electric-energy-result}) with that obtained from direct measurements of irrep probabilities in Sec.\,\ref{subsec:link-irrep-measurement} at $\beta=6.0$. The latter is calculated using the formula
\begin{equation}
    \langle E^2 \rangle = \sum_{R} p(R) C_2(R) ,
\end{equation}
where $p(R)$ is calculated using Eq.\,(\ref{eq:real-probability-using-mc}) and $C_2(R)$ is the Casimir of the irrep $R$. Doing these calculations yields
\begin{equation}
    \langle E^2  \rangle_{\text{PT}} (g=1.0) = 6.37 \quad , \quad \langle E^2 \rangle_{\text{lat}} (g=1.0) \approx 2.5 - 2.7 .
\end{equation}

This is the old and well-known issue of matching couplings between perturbation theory and lattice measurements. There are two sources: 1. Requirement of anisotropy renormalization~\cite{hasenfratz-hamiltonian-scale,anisotropy-renormalization}, and 2. UV contributions of high momentum modes of order $\sim \pi/a$ to lattice observables~\cite{Lepage-tadpole}. The anisotropy renormalization factors in the Hamiltonian limit have been determined for leading order in $\beta$ and are of order $\sim 0.2$~\cite{anisotropy-renormalization}. These small corrections therefore contribute relatively minimally to  the discrepancy between lattice measurements and perturbation theory.

A bigger source of error are the so called tadpole corrections, which can be accounted by dressing the bare coupling $g$ with proxies for high momentum corrections such as the local plaquette expectation value. With tadpole corrections, the lattice action is given by~\cite{tadpole-improvement-hart-hogan}
\begin{equation}
    S = \frac{\beta_{\text{TI}}}{3} \left[ \frac{1}{\xi_{\text{TI}}} \sum_P \left(N - \frac{\Retr  U_P}{u_s^4} \right) + \xi_{\text{TI}} \sum_{\ell,t} \left(N - \frac{\Retr V_\ell (t)}{u_s^2 u_t^2} \right) \right],
\end{equation}
where $\beta$ and $\xi$ are the tadpole corrected coupling and anisotropies, and $u_s$ and $u_t$ can be calculated by using the measured plaquette expectation values such that
\begin{equation}
    u_s^4 = \Big\langle \frac{1}{3} \Retr U_P \Big\rangle \quad ; \quad u_s^2 u_t^2 = \Big\langle \frac{1}{3} \Retr V_l(t) \Big\rangle.
\end{equation}
The tadpole corrected coupling and anisotropy $\beta_{\text{TI}}$ and $\xi_{\text{TI}}$ used in lattice measurements can be related to the corresponding bare quantities by a simple rescaling given by
\begin{subequations}
    \begin{align}
        &\beta_{\text{TI}} = \beta \cdot (u_s^3u_t), \\
        &\xi_{\text{TI}} = \xi \cdot \frac{u_t}{u_s}.
    \end{align}
\end{subequations}

The tadpole corrected coupling $\beta_{\text{TI}}$ can be thought of as consisting of re-summed higher-order corrections due to small-wavelength fluctuations of the gauge fields. Using such an improved expansion parameter $g_{\text{TI}}^2 = 6/\beta_{\text{TI}}$ in lattice perturbation theory improves agreement of $\langle E^2 \rangle$ calculated by Monte-Carlo, as shown in Table.\,\ref{tab:tadpole_combined}. 

\begin{table}[ht!]
  \centering
  \caption{Expectation value of link electric energy $\langle E^2 \rangle$ obtained from tadpole improved lattice perturbation theory compared with Monte Carlo (MC) data from Wilson and heat kernel actions. Tadpole factors $u_s^{\text{PT}}$ and $u_t^{\text{PT}}$ are taken from Ref.~\cite{tadpole-improvement-hart-hogan}}
  \label{tab:tadpole_combined}
  \begin{tabular}{ccccccccc}
    \hline\hline
    $\beta$ & $u_s^{\rm MC}$ & $u_s^{\rm PT}$ & $u_t^{\rm MC}$ & $u_t^{\rm PT}$ & $\beta_{\text{TI}}$ & $\langle E^2\rangle_{\rm TI}$ & $\langle E^2\rangle_{\rm MC}$ \\
    \hline
    \multicolumn{8}{c}{\textit{Wilson action}} \\
    \hline
    6.000 & 0.80765(1) & 0.86194 & 1.23468(2) & 1.00000 & 3.9028(2) & 4.1422(2) & 2.72(5) \\
    6.247 & 0.81896(2) & 0.86806 & 1.21676(3) & 1.00000 & 4.1750(3) & 4.4311(3) & 3.35(11) \\
    6.510 & 0.83009(2) & 0.87401 & 1.20088(3) & 1.00000 & 4.4716(3) & 4.7459(3) & 3.69(9) \\
    6.790 & 0.84041(1) & 0.87979 & 1.18616(2) & 1.00000 & 4.7806(2) & 5.0738(2) & 3.67(8) \\
    7.089 & 0.84939(1) & 0.88541 & 1.17286(2) & 1.00000 & 5.0951(2) & 5.4076(2) & 4.26(6) \\
    \hline
    \multicolumn{8}{c}{\textit{Heat Kernel action}} \\
    \hline
    6.000 & 0.81833(3) & 0.86194 & 1.16387(5) & 1.00000 & 3.8268(5) & 4.0615(5) & 2.57(6) \\
    6.247 & 0.82986(1) & 0.86806 & 1.15318(2) & 1.00000 & 4.1171(2) & 4.3696(2) & 3.24(12) \\
    6.510 & 0.84003(1) & 0.87401 & 1.14004(1) & 1.00000 & 4.3994(1) & 4.6692(1) & 3.57(10) \\
    6.790 & 0.84963(1) & 0.87979 & 1.12756(2) & 1.00000 & 4.6956(2) & 4.9836(2) & 3.89(8) \\
    7.089 & 0.85848(2) & 0.88541 & 1.12055(2) & 1.00000 & 5.0259(3) & 5.3342(3) & 4.08(7) \\
    \hline\hline
  \end{tabular}
\end{table}

Despite improved agreement after tadpole corrections, a significant discrepancy between the lattice measurements and leading-order perturbation theory persists. Evidently higher orders in perturbation theory are needed to reach agreement, which may reflect a need to run the coupling to a somewhat lower scale to match the observable. Further analysis is needed to resolve this difference.

\section{Conclusions} \label{sec:conclusions}
Using perturbative and Monte Carlo methods we have estimated electric properties of the pure $SU(3)$ KS ground state, including the electric energy density, link irrep distribution, and site-singlet population. We estimated the magnitude of local fluctuations in electric energy, which are naturally bounded in a lattice regularization scheme, using lattice perturbation theory in the asymptotic limit $g \to0$ and established the convergence of long-distance electric correlators on the lattice to continuum perturbation theory in Sec.\,\ref{sec:weak-pt}. The UV structure of a discretized theory at finite $g$ can therefore be faithfully represented by a collection of low-lying $SU(3)$ irreps without affecting long-wavelength physics spread over multiple lattice sites. Secs.\,\ref{sec:seamed-partition-function}, \ref{sec:mc_implementation} focused on developing Monte Carlo methods to probe the non-perturbative electric properties of pure $SU(3)$ at practically relevant couplings  $\beta\sim6.0$. The challenge of measuring electric basis observables in the path integral formalism, which employs a group-element measure, was overcome by the introduction of seams on the measured links. Numerical results of Monte Carlo sampling of the Wilson and heat kernel actions for varying degrees of anisotropies and extrapolation to $\xi \to \infty$ was presented in Sec.\,\ref{sec:numerical_results}. At $\beta=6.0$ the four index tensor with $\text{dim} \, R = 27$ was highly suppressed as compared to the lower index tensors, suggesting a truncation threshold $p+q \leq 3$ of the KS Hilbert space for faithful reproduction of low-energy continuum physics.

Several avenues for future exploration naturally arise from the results presented in Sec\,\ref{sec:numerical_results}. The signal-to-noise ratio for site-singlet measurements can be  improved by a bigger Markov chain sample space. Cost-effective generation of decorrelated and ergodic samples require improved methods such as flow-based generative models~\cite{Albergo2019,Kanwar2020}, which train a normalizing flow to directly sample from an approximation to the target distribution, yielding configurations that are statistically independent and can be corrected to the exact distribution via importance reweighting or a Metropolis--Hastings accept/reject step. Finer resolution of site-singlet population can be used to construct improved truncation schemes by removing unimportant site-singlets from the KS Hilbert space which might cut down on the resources required to study Yang-Mills phenomenology on currently available NISQ devices. A program of improved path-integral sampling could also be used to measure other electric observables such as long distance electric correlators and electric flux between static charges. 

We have focused on pure gauge theory in $d=3$ spatial dimensions at small volume. Since the quantities of interest are UV-dominated the finite volume effects are small, but the other properties would be interesting to generalize; for example, early fault tolerant simulations of nonabelian lattice gauge theories will focus on $d=2$ spatial dimensions. Likewise exploratory studies to probe important short-wavelength contributions to the KS vacuum in the presence of a few light dynamical fermions could inform resource estimates for quantum simulations of QCD.

\section*{Aknowledgements}
 This work was supported by the U.S. Department of Energy, Office of Science, Office of High Energy Physics Quantum Information
Science Enabled Discovery (QuantISED) program. We acknowledge the use of the Delta cluster supported by the National Science Foundation [award OAC 2005572] at National Center for Supercomputing Applications through allocation PHY250123 from the Advanced Cyberinfrastructure Coordination Ecosystem: Services \& Support (ACCESS) program, which is supported by U.S. National Science Foundation grants \#2138259, \#2138286, \#2138307, \#2137603, and \#2138296. 
This work utilized LLMs for coding assistance. All results are the responsibility of the authors. We thank Enrico Rinaldi for discussions.

\appendix

\section{Reduction of Lattice Electric Correlators to Bessel Functions}
\label{app:bessel}

The electric correlator integrals over the three-dimensional Brillouin zone
can be reduced to one-dimensional integrals over a Schwinger parameter
involving modified Bessel functions of the first kind $I_n$.

The Schwinger parametrization of $1/\hat{p}$ is given as
\begin{equation}\label{eq:schwinger}
    \frac{1}{|\hat{p}|}
    = \frac{1}{\sqrt{\hat{p}^2}}
    = \frac{1}{\sqrt{\pi}}\int_0^\infty \frac{dt}{\sqrt{t}}\,e^{-t\hat{p}^2},
\end{equation}
which follows from the standard Laplace-transform identity
$1/\sqrt{x} = \pi^{-1/2}\int_0^\infty t^{-1/2}e^{-tx}\,dt$ ($x>0$). When inserted into the integrals in Eqs.\,(\ref{eq:g11-e1}), (\ref{eq:g11-e12}) and (\ref{eq:g12-e12}), the expressions factorize into one-dimensional integrals of the form
\begin{equation}
    \int_{-\pi}^{\pi}\frac{dp}{2\pi}\,e^{inp}\,e^{2t\cos p} = I_n(2t),
    \label{eq:bessel-gen}
\end{equation}
where $I_n$ is the modified Bessel function of the first kind of order $n$.
This follows directly from the generating function
$e^{t(\xi+\xi^{-1})} = \sum_{n=-\infty}^{\infty}\xi^n I_n(2t)$
upon setting $\xi = e^{ip}$. Using Eq.\,\eqref{eq:bessel-gen} together with $\cos p = \tfrac{1}{2}(e^{ip}+e^{-ip})$:

\begin{subequations}
\begin{align}
    \mathcal{A}_n(t)
    &\equiv \int_{-\pi}^{\pi}\frac{dp}{2\pi}\,e^{inp}\,2(1-\cos p)\,e^{2t\cos p}
     = 2I_n(2t) - \bigl[I_{n+1}(2t)+I_{n-1}(2t)\bigr],
    \label{eq:aux-A}\\
    \mathcal{B}_n(t)
    &\equiv \int_{-\pi}^{\pi}\frac{dp}{2\pi}\,e^{inp}(e^{ip}-1)\,e^{2t\cos p}
     = I_{n+1}(2t) - I_n(2t),
    \label{eq:aux-B}\\
    \mathcal{C}_n(t)
    &\equiv \int_{-\pi}^{\pi}\frac{dp}{2\pi}\,e^{inp}(1-e^{-ip})\,e^{2t\cos p}
     = I_n(2t) - I_{n-1}(2t).
    \label{eq:aux-C}
\end{align}
\end{subequations}

For $n=0$, Eq.\,\eqref{eq:aux-A} reduces to
$\mathcal{A}_0 = 2\bigl[I_0(2t)-I_1(2t)\bigr]$
using the symmetry $I_{-1}(x) = I_1(x)$. In terms of $\mathcal{A}_n$, $\mathcal{B}_n$ and $\mathcal{C}_n$, the expressions for the electric correlators are given as
\begin{subequations}
    \begin{align}
        &G_{11}(na \hat{e}_1) = \frac{1}{a^4 \sqrt{\pi}} \int_0^\infty \frac{dt}{\sqrt{t}} \ e^{-6t} \ I_n(2t)  I_0(2t) \mathcal{A}_n(2t), \\
        &G_{11}(na\hat{e}_1 + na \hat{e}_2) = \frac{1}{2a^4 \sqrt{\pi}} \int_0^\infty \frac{dt}{\sqrt{t}} \ e^{-6t} \left[I_0(2t)\mathcal{A}_n(2t)I_n(2t) + I_n^2(2t) \mathcal{A}_0(2t)  \right], \\
        &G_{12}(na\hat{e}_1 + na\hat{e}_2) = \frac{1}{2a^4 \sqrt{\pi}} \int_0^\infty \frac{dt}{\sqrt{t}} \ e^{6t} \ I_0(2t) \mathcal{B}_n(t) \mathcal{C}_n(t).
    \end{align}
\end{subequations}
Substituting the values of $\mathcal{A}$, $\mathcal{B}$ and $\mathcal{C}$ gives the expressions given in the main text.

\section{Determining GT basis states for $SU(3)$ irreps}\label{app:gt-basis}
In this section, we will describe the procedure of finding Gelfand-Tsetlin (GT) basis states for irreps of $SU(3)$ by diagonalizing combinations of the Gell-Mann matrices. All definitions used have been adapted from Ref.\,\cite{Alex2011}.

The basis states of any irrep $R$ of $SU(N)$ can be labelled by ``Gelfand-Tsetlin" (GT) patterns, which is a scheme that labels basis states by iteratively decomposing the carrier space of a $su(K)$ algebra into a direct sum of $su(K-1)$ sub-spaces. In particular, the GT-pattern for a basis state $\ket{M}$ of the $SU(3)$ irrep $R$ is given as
\begin{equation}
\label{eq:gt-pattern}
M = 
\left(
{\setlength{\arraycolsep}{1pt}
\begin{array}{ccccc}
  m_{1,3} && m_{2,3} && m_{3,3} \\
  & m_{1,2} && m_{2,2} & \\
  && m_{1,1} &&
\end{array}}
\right).
\end{equation}
Here, the first row is the ``irrep weight" (or i-weight) of the irrep $R$. By convention, $m_{3,3} = 0$. The second and third rows are the i-weights of the $su(2)$ and $su(1)$ subspaces under which $\ket{M}$ transforms. Indices of the GT-pattern follow the betweenness condition given by
\begin{equation}
\label{eq:betweenness}
m_{k,l} \ge m_{k,l-1} \ge m_{k+1,l}, \qquad 1 \le k < l \le N.
\end{equation}

We can map these GT-patterns to the canonical roots and weights description of basis states by expressing the eigenvalues (or the ``z-weights") of the Cartan generators $J_z^{(1)}$ and $J_z^{(2)}$ in terms of indices of the GT-pattern using the equations
\begin{subequations}\label{eq:z-weight-to-gt-pattern}
    \begin{align}
       & \lambda_1^{M} = m_{1,1} - \frac{1}{2} \left(m_{1,2} + m_{2,2} \right), \label{eq:z-weight-1} \\
       & \lambda_2^M = m_{1,2} + m_{2,2} - \frac{1}{2} \left(m_{1,3} + m_{2,3} \right), \label{eq:z-weight-2}
    \end{align}
\end{subequations}
where $J_z^{(1)}\ket{M} =\lambda_1^M \ket{M}$ and $J_z^{(2)}\ket{M} =\lambda_2^M \ket{M}$. 

The goal in the main text is to find the GT basis given the $SU(3)$ generator matrices $\{T_a , 1 \leq a\leq8\}$ expressed in an arbitrary basis of the representation $R$ with the known $SU(3)$ i-weight $(m_{1,3},m_{2,3},0)$. We can do this by diagonalizing $J_z^{(1)}$ and $J_z^{(2)}$ which can be written as a combination $T_1$ and $T_8$:
\begin{subequations}
    \begin{align}
        &J_z^{(1)} = T_3, \\
        & J_z^{(2)} = - \frac{1}{2} T_3 + \frac{\sqrt{3}}{2} T_8.
    \end{align}
\end{subequations}
Diagonalizing the two Cartan generators gives us the GT basis states $\ket{M}$ and their corresponding z-weights $\lambda_1^M$ and $\lambda_2^M$. We can solve for GT-pattern indices $m_{1,1}$ and $m_{1,1} +m_{1,2}$ from values of z-weights using Eq.\,(\ref{eq:z-weight-to-gt-pattern}). However, since the z-weights only render two equations, we cannot determine the values of $m_{1,1}$ and $m_{1,2}$ exactly.

In order to determine the exact split, we rely on the fact that the Gell-Mann matrices $T_1$, $T_2$, $T_3$ form a $su(2)$ sub-algebra and therefore have representations that can be labeled by the isospin operator 
\begin{equation}
    I = T_1^2 + T_2^2 + T_3^2.
\end{equation}
We can relate the eigenvalues $I^M$ of this isospin operator to indices of the GT pattern of $\ket{M}$ by using the betweenness condition $m_{1,2} \leq m_{1,1} \leq m_{2,2}$ and the $T_3$ eigenvalue in Eq.\,(\ref{eq:z-weight-1})  to obtain the inequality
\begin{equation}
    \frac{m_{1,2} -m_{1,1}}{2} \leq \lambda_1^M \leq \frac{m_{1,1} - m_{1,2}}{2},
\end{equation}
which means that $m_{1,1}-m_{1,2} = 2I^M$. Therefore, a GT-basis state can be unambiguously determined by the three eigenvalues $\lambda_1^M$, $\lambda_2^M$ and $I^M$. So, to find the GT-basis states, one should simultaneously diagonalize $J_z^{(1)}$, $J_z^{(2)}$ and $I$.

\bibliographystyle{unsrtnat}
\bibliography{refs}

\begin{thebibliography}{52}
\providecommand{\natexlab}[1]{#1}
\providecommand{\url}[1]{\texttt{#1}}
\expandafter\ifx\csname urlstyle\endcsname\relax
  \providecommand{\doi}[1]{doi: #1}\else
  \providecommand{\doi}{doi: \begingroup \urlstyle{rm}\Url}\fi

\bibitem[Feynman(1982)]{Feynman1982}
Richard~P. Feynman.
\newblock Simulating physics with computers.
\newblock \emph{International Journal of Theoretical Physics}, 21\penalty0 (6):\penalty0 467--488, 1982.
\newblock \doi{10.1007/BF02650179}.

\bibitem[Jordan et~al.(2014)Jordan, Lee, and Preskill]{Jordan2012}
Stephen~P. Jordan, Keith S.~M. Lee, and John Preskill.
\newblock Quantum computation of scattering in scalar quantum field theories.
\newblock \emph{Quantum Info. Comput.}, 14\penalty0 (11–12):\penalty0 1014–1080, September 2014.
\newblock ISSN 1533-7146.

\bibitem[Preskill(2018)]{Preskill2018}
John Preskill.
\newblock Quantum {C}omputing in the {NISQ} era and beyond.
\newblock \emph{{Quantum}}, 2:\penalty0 79, August 2018.
\newblock ISSN 2521-327X.
\newblock \doi{10.22331/q-2018-08-06-79}.
\newblock URL \url{https://doi.org/10.22331/q-2018-08-06-79}.

\bibitem[Creutz et~al.(1983)Creutz, Jacobs, and Rebbi]{CREUTZ-LGT-MC-REVIEW}
Michael Creutz, Laurence Jacobs, and Claudio Rebbi.
\newblock Monte carlo computations in lattice gauge theories.
\newblock \emph{Physics Reports}, 95\penalty0 (4):\penalty0 201--282, 1983.
\newblock ISSN 0370-1573.
\newblock \doi{https://doi.org/10.1016/0370-1573(83)90016-9}.
\newblock URL \url{https://www.sciencedirect.com/science/article/pii/0370157383900169}.

\bibitem[Bazavov et~al.(2018)Bazavov, Bernard, Brown, DeTar, El-Khadra, G\'amiz, Gottlieb, Heller, Komijani, Kronfeld, Laiho, Mackenzie, Neil, Simone, Sugar, Toussaint, and Van~de Water]{B-D-DECAY-HYPER-PRECISION}
A.~Bazavov, C.~Bernard, N.~Brown, C.~DeTar, A.~X. El-Khadra, E.~G\'amiz, Steven Gottlieb, U.~M. Heller, J.~Komijani, A.~S. Kronfeld, J.~Laiho, P.~B. Mackenzie, E.~T. Neil, J.~N. Simone, R.~L. Sugar, D.~Toussaint, and R.~S. Van~de Water.
\newblock $b$- and $d$-meson leptonic decay constants from four-flavor lattice qcd.
\newblock \emph{Phys. Rev. D}, 98:\penalty0 074512, Oct 2018.
\newblock \doi{10.1103/PhysRevD.98.074512}.
\newblock URL \url{https://link.aps.org/doi/10.1103/PhysRevD.98.074512}.

\bibitem[Aoki et~al.(2026)Aoki, Blum, Collins, Del~Debbio, Della~Morte, Dimopoulos, Feng, Golterman, Gottlieb, Gupta, Herdo\'{\i}za, Hernandez, J\"uttner, Kaneko, Lunghi, Meinel, Monahan, Nicholson, Onogi, Petreczky, Portelli, Ramos, Sharpe, Simone, Sint, Sommer, Tantalo, Van~de Water, Vaquero, Wenger, and Wittig]{flag-review-2024}
Y.~Aoki, T.~Blum, S.~Collins, L.~Del~Debbio, M.~Della~Morte, P.~Dimopoulos, X.~Feng, M.~Golterman, Steven Gottlieb, R.~Gupta, G.~Herdo\'{\i}za, P.~Hernandez, A.~J\"uttner, T.~Kaneko, E.~Lunghi, S.~Meinel, C.~Monahan, A.~Nicholson, T.~Onogi, P.~Petreczky, A.~Portelli, A.~Ramos, S.~R. Sharpe, J.~N. Simone, S.~Sint, R.~Sommer, N.~Tantalo, R.~Van~de Water, A.~Vaquero, U.~Wenger, and H.~Wittig.
\newblock Flag review 2024.
\newblock \emph{Phys. Rev. D}, 113:\penalty0 014508, Jan 2026.
\newblock \doi{10.1103/nfzp-p5dn}.
\newblock URL \url{https://link.aps.org/doi/10.1103/nfzp-p5dn}.

\bibitem[Borsanyi et~al.(2016)Borsanyi, Fodor, Guenther, Kampert, Katz, Kawanai, Kovacs, Mages, Pasztor, Szabo, Toth, and Szabo]{Borsanyi2016}
Szabolcs Borsanyi, Zoltan Fodor, Jana Guenther, Karl-Heinz Kampert, S{\'a}ndor~D. Katz, Taichi Kawanai, Thomas~G. Kovacs, Simon Mages, Attila Pasztor, Kalman~K. Szabo, B{\'a}lint~C. Toth, and Kalman~K. Szabo.
\newblock Calculation of the axion mass based on high-temperature lattice quantum chromodynamics.
\newblock \emph{Nature}, 539\penalty0 (7627):\penalty0 69--71, 2016.
\newblock \doi{10.1038/nature20115}.

\bibitem[Gavai et~al.(2025)Gavai, Jaensch, Kaczmarek, Karsch, Sarkar, Shanker, Sharma, Sharma, and Ueding]{Finite-Temp-LQCD-CHiral-Crossover}
Rajiv~V. Gavai, Mischa~E. Jaensch, Olaf Kaczmarek, Frithjof Karsch, Mugdha Sarkar, Ravi Shanker, Sayantan Sharma, Sipaz Sharma, and Tristan Ueding.
\newblock Aspects of the chiral crossover transition in ($2+1$)-flavor qcd with m\"obius domain-wall fermions.
\newblock \emph{Phys. Rev. D}, 111:\penalty0 034507, Feb 2025.
\newblock \doi{10.1103/PhysRevD.111.034507}.
\newblock URL \url{https://link.aps.org/doi/10.1103/PhysRevD.111.034507}.

\bibitem[Troyer and Wiese(2005)]{Troyer2005}
Matthias Troyer and Uwe-Jens Wiese.
\newblock Computational complexity and fundamental limitations to fermionic quantum monte carlo simulations.
\newblock \emph{Phys. Rev. Lett.}, 94:\penalty0 170201, May 2005.
\newblock \doi{10.1103/PhysRevLett.94.170201}.
\newblock URL \url{https://link.aps.org/doi/10.1103/PhysRevLett.94.170201}.

\bibitem[Bauer et~al.(2023)Bauer, Davoudi, Balantekin, Bhattacharya, Carena, de~Jong, Draper, El-Khadra, Gemelke, Hanada, Kharzeev, Lamm, Li, Liu, Lukin, Meurice, Monroe, Nachman, Pagano, Preskill, Rinaldi, Roggero, Santiago, Savage, Siddiqi, Siopsis, Van~Zanten, Wiebe, Yamauchi, Yeter-Aydeniz, and Zorzetti]{Bauer2023}
Christian~W. Bauer, Zohreh Davoudi, A.~Baha Balantekin, Tanmoy Bhattacharya, Marcela Carena, Wibe~A. de~Jong, Patrick Draper, Aida El-Khadra, Nate Gemelke, Masanori Hanada, Dmitri Kharzeev, Henry Lamm, Ying-Ying Li, Junyu Liu, Mikhail Lukin, Yannick Meurice, Christopher Monroe, Benjamin Nachman, Guido Pagano, John Preskill, Enrico Rinaldi, Alessandro Roggero, David~I. Santiago, Martin~J. Savage, Irfan Siddiqi, George Siopsis, David Van~Zanten, Nathan Wiebe, Yukari Yamauchi, K\"ubra Yeter-Aydeniz, and Silvia Zorzetti.
\newblock Quantum simulation for high-energy physics.
\newblock \emph{PRX Quantum}, 4:\penalty0 027001, May 2023.
\newblock \doi{10.1103/PRXQuantum.4.027001}.
\newblock URL \url{https://link.aps.org/doi/10.1103/PRXQuantum.4.027001}.

\bibitem[Davoudi et~al.(2020)Davoudi, Hafezi, Monroe, Pagano, Seif, and Shaw]{Davoudi2020}
Zohreh Davoudi, Mohammad Hafezi, Christopher Monroe, Guido Pagano, Alireza Seif, and Andrew Shaw.
\newblock Towards analog quantum simulations of lattice gauge theories with trapped ions.
\newblock \emph{Phys. Rev. Res.}, 2:\penalty0 023015, Apr 2020.
\newblock \doi{10.1103/PhysRevResearch.2.023015}.
\newblock URL \url{https://link.aps.org/doi/10.1103/PhysRevResearch.2.023015}.

\bibitem[Klco et~al.(2022)Klco, Roggero, and Savage]{Klco2022}
N.~Klco, A.~Roggero, and M.~J. Savage.
\newblock Standard model physics and the digital quantum revolution: thoughts about the interface.
\newblock \emph{Reports on Progress in Physics}, 85\penalty0 (6):\penalty0 064301, 2022.
\newblock \doi{10.1088/1361-6633/ac58a4}.

\bibitem[Halimeh et~al.(2025)Halimeh, Mueller, Knolle, Papić, and Davoudi]{halimeh2025quantumsimulationoutofequilibriumdynamics}
Jad~C. Halimeh, Niklas Mueller, Johannes Knolle, Zlatko Papić, and Zohreh Davoudi.
\newblock Quantum simulation of out-of-equilibrium dynamics in gauge theories, 2025.
\newblock URL \url{https://arxiv.org/abs/2509.03586}.

\bibitem[Kogut and Susskind(1975)]{Kogut1975}
John Kogut and Leonard Susskind.
\newblock Hamiltonian formulation of wilson's lattice gauge theories.
\newblock \emph{Phys. Rev. D}, 11:\penalty0 395--408, Jan 1975.
\newblock \doi{10.1103/PhysRevD.11.395}.
\newblock URL \url{https://link.aps.org/doi/10.1103/PhysRevD.11.395}.

\bibitem[Somma et~al.(2003)Somma, Ortiz, Knill, and Gubernatis]{Somma2003}
Rolando~D. Somma, Gerardo Ortiz, Emanuel~H. Knill, and James Gubernatis.
\newblock Quantum simulations of physics problems.
\newblock In Eric Donkor, Andrew~R. Pirich, and Howard~E. Brandt, editors, \emph{Quantum Information and Computation}, volume 5105, page~96. SPIE, August 2003.
\newblock \doi{10.1117/12.487249}.
\newblock URL \url{http://dx.doi.org/10.1117/12.487249}.

\bibitem[Byrnes and Yamamoto(2006)]{Byrnes2006}
Tim Byrnes and Yoshihisa Yamamoto.
\newblock Simulating lattice gauge theories on a quantum computer.
\newblock \emph{Phys. Rev. A}, 73:\penalty0 022328, Feb 2006.
\newblock \doi{10.1103/PhysRevA.73.022328}.
\newblock URL \url{https://link.aps.org/doi/10.1103/PhysRevA.73.022328}.

\bibitem[Siew et~al.(2026)Siew, Chandrasekharan, and Bhattacharya]{siew2026continuumlimitqubitregularizedsu3}
Rui~Xian Siew, Shailesh Chandrasekharan, and Tanmoy Bhattacharya.
\newblock Continuum limit of a qubit-regularized su(3) lattice gauge theory with glueballs, 2026.
\newblock URL \url{https://arxiv.org/abs/2603.01215}.

\bibitem[Jakobs et~al.(2025)Jakobs, Garofalo, Hartung, Jansen, Ostmeyer, Romiti, and Urbach]{Jakobs2025}
Timo Jakobs, Marco Garofalo, Tobias Hartung, Karl Jansen, Johann Ostmeyer, Simone Romiti, and Carsten Urbach.
\newblock Dynamics in hamiltonian lattice gauge theory: approaching the continuum limit with partitionings of su(2).
\newblock \emph{The European Physical Journal C}, 85\penalty0 (4):\penalty0 354, 2025.
\newblock \doi{10.1140/epjc/s10052-025-15120-x}.

\bibitem[Kane et~al.(2025)Kane, Hariprakash, and Bauer]{kane2025obtainingcontinuumphysicsdynamical}
Christopher~F. Kane, Siddharth Hariprakash, and Christian~W. Bauer.
\newblock Obtaining continuum physics from dynamical simulations of hamiltonian lattice gauge theories, 2025.
\newblock URL \url{https://arxiv.org/abs/2506.16559}.

\bibitem[Ciavarella et~al.(2021)Ciavarella, Klco, and Savage]{Ciavarella2021}
Anthony Ciavarella, Natalie Klco, and Martin~J. Savage.
\newblock Trailhead for quantum simulation of su(3) yang-mills lattice gauge theory in the local multiplet basis.
\newblock \emph{Phys. Rev. D}, 103:\penalty0 094501, May 2021.
\newblock \doi{10.1103/PhysRevD.103.094501}.
\newblock URL \url{https://link.aps.org/doi/10.1103/PhysRevD.103.094501}.

\bibitem[Ciavarella and Bauer(2024)]{Ciavarella-large-N-truncation}
Anthony~N. Ciavarella and Christian~W. Bauer.
\newblock Quantum simulation of su(3) lattice yang-mills theory at leading order in large-${N}_{c}$ expansion.
\newblock \emph{Phys. Rev. Lett.}, 133:\penalty0 111901, Sep 2024.
\newblock \doi{10.1103/PhysRevLett.133.111901}.
\newblock URL \url{https://link.aps.org/doi/10.1103/PhysRevLett.133.111901}.

\bibitem[Alexandru et~al.(2022)Alexandru, Bedaque, Brett, and Lamm]{Alexandru-S1080-truncation}
Andrei Alexandru, Paulo~F. Bedaque, Ruair\'{\i} Brett, and Henry Lamm.
\newblock Spectrum of digitized qcd: Glueballs in a $s(1080)$ gauge theory.
\newblock \emph{Phys. Rev. D}, 105:\penalty0 114508, Jun 2022.
\newblock \doi{10.1103/PhysRevD.105.114508}.
\newblock URL \url{https://link.aps.org/doi/10.1103/PhysRevD.105.114508}.

\bibitem[Gonz{\'a}lez-Cuadra et~al.(2025)Gonz{\'a}lez-Cuadra, Wang, Keesling, et~al.]{GonzalezCuadra2025}
Daniel Gonz{\'a}lez-Cuadra, Y.~T. Wang, Alexander Keesling, et~al.
\newblock Observation of string breaking on a (2+1)d rydberg quantum simulator.
\newblock \emph{Nature}, 642\penalty0 (8067):\penalty0 321--326, 2025.
\newblock \doi{10.1038/s41586-025-09051-6}.

\bibitem[Xu et~al.(2026)Xu, Borla, Hemery, Joshi, Dreyer, Rinaldi, and Halimeh]{xu2026observationglueballexcitationsstring}
Kaidi Xu, Umberto Borla, Kevin Hemery, Rohan Joshi, Henrik Dreyer, Enrico Rinaldi, and Jad~C. Halimeh.
\newblock Observation of glueball excitations and string breaking in a $2+1$d $\mathbb{Z}_2$ lattice gauge theory on a trapped-ion quantum computer, 2026.
\newblock URL \url{https://arxiv.org/abs/2604.07435}.

\bibitem[Froland et~al.(2025)Froland, Grabowska, and Li]{froland2025simulatingfullygaugefixedsu2}
Henry Froland, Dorota~M. Grabowska, and Zhiyao Li.
\newblock Simulating fully gauge-fixed su(2) hamiltonian dynamics on digital quantum computers, 2025.
\newblock URL \url{https://arxiv.org/abs/2512.22782}.

\bibitem[Ciavarella et~al.(2026)Ciavarella, de~Putter, Younis, and Rrapaj]{ciavarella2026quantumsimulationstwodimensionalnonabelian}
Anthony~N. Ciavarella, Roland de~Putter, Ed~Younis, and Ermal Rrapaj.
\newblock Quantum simulations of two-dimensional non-abelian adjoint string breaking, 2026.
\newblock URL \url{https://arxiv.org/abs/2608.28752}.

\bibitem[Chernyshev et~al.(2026)Chernyshev, Farrell, Illa, Savage, Maksymov, Tripier, Lopez-Ruiz, Arrasmith, de~Sereville, Brodutch, Girotto, Kaushik, and Roetteler]{Chernyshev2026}
Ivan~A. Chernyshev, Roland~C. Farrell, Marc Illa, Martin~J. Savage, Andrii Maksymov, Felix Tripier, Miguel~Angel Lopez-Ruiz, Andrew Arrasmith, Yvette de~Sereville, Aharon Brodutch, Claudio Girotto, Ananth Kaushik, and Martin Roetteler.
\newblock Pathfinding quantum simulations of neutrinoless double-$\beta$ decay.
\newblock \emph{Nature Communications}, 17\penalty0 (1):\penalty0 812, 2026.
\newblock \doi{10.1038/s41467-026-68536-8}.

\bibitem[Balaji et~al.(2026{\natexlab{a}})Balaji, Draper, and Rinaldi]{balaji2026timedomainobservablesresonantscattering}
Praveen Balaji, Patrick Draper, and Enrico Rinaldi.
\newblock Time-domain observables for resonant scattering in quantum simulations of lattice gauge theory, 2026{\natexlab{a}}.
\newblock URL \url{https://arxiv.org/abs/2610.17470}.

\bibitem[Balaji et~al.(2025)Balaji, Conefrey-Shinozaki, Draper, Elhaderi, Gupta, Hidalgo, Lytle, and Rinaldi]{quantum-circuits-draper}
Praveen Balaji, Cian\'an Conefrey-Shinozaki, Patrick Draper, Jason~K. Elhaderi, Drishti Gupta, Luis Hidalgo, Andrew Lytle, and Enrico Rinaldi.
\newblock Quantum circuits for su(3) lattice gauge theory.
\newblock \emph{Phys. Rev. D}, 112:\penalty0 054511, Sep 2025.
\newblock \doi{10.1103/k8f6-yft8}.
\newblock URL \url{https://link.aps.org/doi/10.1103/k8f6-yft8}.

\bibitem[Balaji et~al.(2026{\natexlab{b}})Balaji, Conefrey-Shinozaki, Draper, Elhaderi, Gupta, Hidalgo, and Lytle]{B-truncation}
Praveen Balaji, Cian\'an Conefrey-Shinozaki, Patrick Draper, Jason~K. Elhaderi, Drishti Gupta, Luis Hidalgo, and Andrew Lytle.
\newblock Perturbation theory, irrep truncations, and state preparation methods for quantum simulations of $su(3)$ lattice gauge theory.
\newblock \emph{Phys. Rev. D}, 113:\penalty0 094505, May 2026{\natexlab{b}}.
\newblock \doi{10.1103/m719-7tdf}.
\newblock URL \url{https://link.aps.org/doi/10.1103/m719-7tdf}.

\bibitem[Lamm et~al.(2024)Lamm, Li, Shu, Wang, and Xu]{Lamm2024}
Henry Lamm, Ying-Ying Li, Jing Shu, Yi-Lin Wang, and Bin Xu.
\newblock Block encodings of discrete subgroups on a quantum computer.
\newblock \emph{Phys. Rev. D}, 110:\penalty0 054505, Sep 2024.
\newblock \doi{10.1103/PhysRevD.110.054505}.
\newblock URL \url{https://link.aps.org/doi/10.1103/PhysRevD.110.054505}.

\bibitem[Rhodes et~al.(2024)Rhodes, Kreshchuk, and Pathak]{Rhodes2024}
Mason~L. Rhodes, Michael Kreshchuk, and Shivesh Pathak.
\newblock Exponential improvements in the simulation of lattice gauge theories using near-optimal techniques.
\newblock \emph{PRX Quantum}, 5:\penalty0 040347, Dec 2024.
\newblock \doi{10.1103/PRXQuantum.5.040347}.
\newblock URL \url{https://link.aps.org/doi/10.1103/PRXQuantum.5.040347}.

\bibitem[Kan and Nam(2022)]{kan2022latticequantumchromodynamicselectrodynamics}
Angus Kan and Yunseong Nam.
\newblock Lattice quantum chromodynamics and electrodynamics on a universal quantum computer, 2022.
\newblock URL \url{https://arxiv.org/abs/2107.12769}.

\bibitem[Draper(2026)]{draper2026blockencodingnonabelianlattice}
Patrick Draper.
\newblock Block encoding non-abelian lattice gauge theory, 2026.
\newblock URL \url{https://arxiv.org/abs/2608.17115}.

\bibitem[Hidalgo and Draper(2026)]{Hidalgo2026}
Luis Hidalgo and Patrick Draper.
\newblock Lattice quantum chromodynamics for quantum simulations, 2026.
\newblock URL \url{https://arxiv.org/abs/2607.26445}.

\bibitem[Creutz(1977)]{creutz-transfer-matrix}
Michael Creutz.
\newblock Gauge fixing, the transfer matrix, and confinement on a lattice.
\newblock \emph{Phys. Rev. D}, 15:\penalty0 1128--1136, Feb 1977.
\newblock \doi{10.1103/PhysRevD.15.1128}.
\newblock URL \url{https://link.aps.org/doi/10.1103/PhysRevD.15.1128}.

\bibitem[Hasenfratz and Hasenfratz(1981)]{hasenfratz-hamiltonian-scale}
Anna Hasenfratz and Péter Hasenfratz.
\newblock The scales of euclidean and hamiltonian lattice qcd.
\newblock \emph{Nuclear Physics B}, 193\penalty0 (1):\penalty0 210--220, 1981.
\newblock ISSN 0550-3213.
\newblock \doi{https://doi.org/10.1016/0550-3213(81)90526-5}.
\newblock URL \url{https://www.sciencedirect.com/science/article/pii/0550321381905265}.

\bibitem[Karsch(1982)]{Karsch1982}
F.~Karsch.
\newblock {SU(N) Gauge Theory Couplings on Asymmetric Lattices}.
\newblock \emph{Nucl. Phys. B}, 205:\penalty0 285--300, 1982.
\newblock \doi{10.1016/0550-3213(82)90390-X}.

\bibitem[Carleo et~al.(2010)Carleo, Becca, Moroni, and Baroni]{Worm-Operator}
Giuseppe Carleo, Federico Becca, Saverio Moroni, and Stefano Baroni.
\newblock Reptation quantum monte carlo algorithm for lattice hamiltonians with a directed-update scheme.
\newblock \emph{Phys. Rev. E}, 82:\penalty0 046710, Oct 2010.
\newblock \doi{10.1103/PhysRevE.82.046710}.
\newblock URL \url{https://link.aps.org/doi/10.1103/PhysRevE.82.046710}.

\bibitem[Wilson(1974)]{Wilson1974}
Kenneth~G. Wilson.
\newblock Confinement of quarks.
\newblock \emph{Phys. Rev. D}, 10:\penalty0 2445--2459, Oct 1974.
\newblock \doi{10.1103/PhysRevD.10.2445}.
\newblock URL \url{https://link.aps.org/doi/10.1103/PhysRevD.10.2445}.

\bibitem[Menotti and Onofri(1981)]{Menotti1981288}
P.~Menotti and E.~Onofri.
\newblock The action of su(n) lattice gauge theory in terms of the heat kernel on the group manifold.
\newblock 190\penalty0 (2):\penalty0 288 – 300, 1981.
\newblock \doi{10.1016/0550-3213(81)90560-5}.

\bibitem[Drouffe and Zuber(1983)]{Drouffe1983}
Jean-Michel Drouffe and Jean-Bernard Zuber.
\newblock Strong coupling and mean field methods in lattice gauge theories.
\newblock \emph{Physics Reports}, 102\penalty0 (1):\penalty0 1--119, 1983.
\newblock ISSN 0370-1573.
\newblock \doi{https://doi.org/10.1016/0370-1573(83)90034-0}.
\newblock URL \url{https://www.sciencedirect.com/science/article/pii/0370157383900340}.

\bibitem[Lepage and Mackenzie(1993)]{Lepage-tadpole}
G.~Peter Lepage and Paul~B. Mackenzie.
\newblock Viability of lattice perturbation theory.
\newblock \emph{Phys. Rev. D}, 48:\penalty0 2250--2264, Sep 1993.
\newblock \doi{10.1103/PhysRevD.48.2250}.
\newblock URL \url{https://link.aps.org/doi/10.1103/PhysRevD.48.2250}.

\bibitem[Creutz(1980)]{creutz-heatbath}
Michael Creutz.
\newblock Monte carlo study of quantized su(2) gauge theory.
\newblock \emph{Phys. Rev. D}, 21:\penalty0 2308--2315, Apr 1980.
\newblock \doi{10.1103/PhysRevD.21.2308}.
\newblock URL \url{https://link.aps.org/doi/10.1103/PhysRevD.21.2308}.

\bibitem[Cabibbo and Marinari(1982)]{cabibbo-pseudo-heatbath}
Nicola Cabibbo and Enzo Marinari.
\newblock A new method for updating su(n) matrices in computer simulations of gauge theories.
\newblock \emph{Physics Letters B}, 119\penalty0 (4):\penalty0 387--390, 1982.
\newblock ISSN 0370-2693.
\newblock \doi{https://doi.org/10.1016/0370-2693(82)90696-7}.
\newblock URL \url{https://www.sciencedirect.com/science/article/pii/0370269382906967}.

\bibitem[Fulton and Harris(2004)]{fulton1991representation}
William Fulton and Joe Harris.
\newblock \emph{Representation Theory: A First Course}, volume 129 of \emph{Graduate Texts in Mathematics}.
\newblock Springer, New York, NY, 2004.
\newblock ISBN 978-1-4612-0979-9.
\newblock \doi{10.1007/978-1-4612-0979-9}.

\bibitem[{National Institute of Standards and Technology}()]{NIST:EngineeringStatisticsHandbook:3.1}
{National Institute of Standards and Technology}.
\newblock 3.1. how can we determine if two processes have the same mean?
\newblock In \emph{NIST/SEMATECH e-Handbook of Statistical Methods}. National Institute of Standards and Technology, Gaithersburg, MD.
\newblock URL \url{https://www.itl.nist.gov/div898/handbook/prc/section3/prc31.htm}.

\bibitem[Byrnes et~al.(2004)Byrnes, Loan, Hamer, Bonnet, Leinweber, Williams, and Zanotti]{anisotropy-renormalization}
T.~M.~R. Byrnes, M.~Loan, C.~J. Hamer, Fr\'ed\'eric D.~R. Bonnet, Derek~B. Leinweber, Anthony~G. Williams, and James~M. Zanotti.
\newblock Hamiltonian limit of (3+1)-dimensional su(3) lattice gauge theory on anisotropic lattices.
\newblock \emph{Phys. Rev. D}, 69:\penalty0 074509, Apr 2004.
\newblock \doi{10.1103/PhysRevD.69.074509}.
\newblock URL \url{https://link.aps.org/doi/10.1103/PhysRevD.69.074509}.

\bibitem[Hart et~al.(2004)Hart, Horgan, and Storoni]{tadpole-improvement-hart-hogan}
A.~Hart, R.~R. Horgan, and L.~C. Storoni.
\newblock Perturbation theory versus simulation for tadpole improvement factors in pure gauge theories.
\newblock \emph{Phys. Rev. D}, 70:\penalty0 034501, Aug 2004.
\newblock \doi{10.1103/PhysRevD.70.034501}.
\newblock URL \url{https://link.aps.org/doi/10.1103/PhysRevD.70.034501}.

\bibitem[Albergo et~al.(2019)Albergo, Kanwar, and Shanahan]{Albergo2019}
M.~S. Albergo, G.~Kanwar, and P.~E. Shanahan.
\newblock Flow-based generative models for markov chain monte carlo in lattice field theory.
\newblock \emph{Phys. Rev. D}, 100:\penalty0 034515, Aug 2019.
\newblock \doi{10.1103/PhysRevD.100.034515}.
\newblock URL \url{https://link.aps.org/doi/10.1103/PhysRevD.100.034515}.

\bibitem[Kanwar et~al.(2020)Kanwar, Albergo, Boyda, Cranmer, Hackett, Racani{\`e}re, Rezende, and Shanahan]{Kanwar2020}
Gurtej Kanwar, Michael~S. Albergo, Denis Boyda, Kyle Cranmer, Daniel~C. Hackett, S{\'e}bastien Racani{\`e}re, Danilo~Jimenez Rezende, and Phiala~E. Shanahan.
\newblock Equivariant flow-based sampling for lattice gauge theory.
\newblock \emph{Phys. Rev. Lett.}, 125:\penalty0 121601, Sep 2020.
\newblock \doi{10.1103/PhysRevLett.125.121601}.
\newblock URL \url{https://link.aps.org/doi/10.1103/PhysRevLett.125.121601}.

\bibitem[Alex et~al.(2011)Alex, Kalus, Huckleberry, and von Delft]{Alex2011}
Arne Alex, Matthias Kalus, Alan Huckleberry, and Jan von Delft.
\newblock A numerical algorithm for the explicit calculation of {SU($N$)} and {$\mbox{SL}(N,\mathbb {C})$} {Clebsch--Gordan} coefficients.
\newblock \emph{Journal of Mathematical Physics}, 52\penalty0 (2):\penalty0 023507, 2011.
\newblock \doi{10.1063/1.3551525}.

\end{thebibliography}

\end{document}